\documentclass[a4paper]{jpconf}
\usepackage{graphicx}
\usepackage{amssymb}
\usepackage{amsmath}
\usepackage{dcolumn}
\usepackage{bm}

\renewcommand{\vec}[1]{\bm{\mathrm{#1}}}

\begin{document}



\title{Theory for Quartet Condensation in Fermi Systems with Applications to Nuclei and Nuclear Matter}

\author{P.\ Schuck$^1$ $^2$ $^3$,Y.\ Funaki$^4$, H. \ Horiuchi$^5$, G.\ R\"opke$^6$,  A. \ Tohsaki$^5$, T. \ Yamada$^7$} 



\address{$^1$Institut de Physique Nucl\'eaire, CNRS, UMR8608, \\ 

Orsay, F-91406, France\\

$^2$Universit\'e Paris-Sud, Orsay, F-91505, France \\

$^3$Laboratoire de Physique et de Mod\'elisation des Milieux Condens\'es, CNRS et Universit\'e Joseph Fourier, UMR5493, 25 Av. des Martyrs, BP 166, F-38042 Grenoble Cedex 9, France\\

$^4$The Institute of Physical and Chemical Research (RIKEN), Wako, \\

Saitama, 351-0198, Japan \\

$^5$Research Center for Nuclear Physics (RCNP), Osaka University, \\

Ibaraki, Osaka 567-0047, Japan \\

$^6$Institut f\"ur Physik, University of Rostock, Universit\"atsplatz 1, \\

18051 Rostock, Germany \\

$^7$Laboratory of Physics, Kanto Gakuin University, \\

Yokohama 236-8501, Japan

}



\begin{abstract}
The theory of quartet condensation is further developed.
The onset of quartetting in homogeneous fermionic matter is studied with 
the help of an in-medium modified four fermion 
equation. It is found that at very low density quartetting wins over pairing. 
At zero temperature, in analogy to pairing, a set of equations for the quartet order parameter 
is given. Contrary to pairing, quartetting only exists for strong coupling and breaks down for weak coupling. Reasons for this finding are detailed. In an application to nuclear matter, the critical 
temperature for $\alpha$ particle condensation can reach values up to around 
8 MeV. The disappearance of 
$\alpha$-particles with increasing density, i.e. the Mott transition, is 
investigated. In finite nuclei the Hoyle state, that is the ${0_2}^+$ of 
$^{12}$C is identified as an '$\alpha$-particle condensate' state. It is 
conjectured that such states also exist in heavier $n\alpha$-nuclei, like 
$^{16}$O, $^{20}$Ne, etc. The sixth $0^+$ state in $^{16}$O is proposed as an 
analogue to the Hoyle state. The Gross-Pitaevski equation is employed to make 
an estimate of the maximum number of $\alpha$ particles a condensate state 
can contain.
Possible quartet condensation in 
other systems is discussed briefly.\\

\noindent 
Keywords: quartet condensation, nuclear matter, $\alpha$-matter, 
superfluidity, Bose-Einstein
condensation, strongly coupled systems

\end{abstract}


\section{Introduction}

One of the most amazing phenomena in quantum many-particle systems is
the formation of quantum condensates.  At present, the formation of
condensates is of particular interest in strongly coupled fermion
systems in which the crossover from Bardeen-Cooper-Schrieffer (BCS)
pairing to Bose-Einstein condensation (BEC) may be investigated. Among
very different quantum systems such as the electron-hole exciton and 
bi-exciton systems in
excited semiconductors, atoms in traps at extremely low temperatures, etc.,
nuclear matter is especially well suited for the study of correlation 
effects in a quantum liquid. However, more exotic systems like the 
dense hydrogen gas may also feature condensation phenomena of electron-electron,
proton-proton, and electron-proton pairs \cite{Ashcroft} where the strongly 
bound 
hydrogen molecule can be thought of as a candidate for condensation in 
the same context.
Coming to nuclear systems,
neutron matter, nuclear matter, but also finite nuclei are superfluid.
However, at low density, nuclear matter will not cluster into pairs, i.e.
deuterons, but rather into $\alpha$ -particles which are much more stable. Also
heavier clusters, starting with Carbon, may be of importance but are 
presently not considered for condensation phenomena. Therefore, one may ask 
the question whether there exists quartetting, i.e. $\alpha$-particle 
condensation, in nuclei, analogous to nuclear pairing. The only nucleus 
which in its ground state
has a pronounced $\alpha$ -cluster structure is $^8$Be. In section 5 we will
show a figure of $^8$Be in the laboratory frame and in the intrinsic deformed
frame. We will see that $^8$Be is formed out of two almost free $\alpha$ -
particles
roughly 4 fm apart, only weakly overlapping with their surface tails. Actually $^8$Be is 
slightly unstable and the two $\alpha$'s only hold together via the Coulomb 
barrier.
Because of the large distance of the two $\alpha$-particles, the 0$^+$ ground
state of $^8$Be has, in the laboratory frame, a spherical density
distribution whose average is very low: about 1/3 of ordinary saturation
density $\rho_0$. $^8$Be is, therefore, a very large object with an rms 
radius of
about 3.7 fm to be compared with the nuclear systematics of $R = r_0A^{1/3}$ = 
2.44 fm. Definitely $^8$Be is a rather unusual and, in its kind, unique 
nucleus. One may ask the question what happens when one brings a third 
$\alpha$-particle alongside of $^8$Be. We know the answer: the 3$\alpha$ 
system collapses to the ground state of $^{12}$C which is much denser than 
$^8$Be and can not accommodate, with its small radius of 2.4 fm, three more or 
less free $\alpha$-particles barely touching one another. One nevertheless may 
ask the question whether the dilute three $\alpha$ configuration $^8$Be-
$\alpha$, or $\alpha-\alpha-\alpha$, may not form an isomeric or 
excited 
state of $^{12}$C. That such a 
state indeed exists will be one of the main subjects of our considerations.
Once one accepts the idea of the existence of an $\alpha$-gas state in 
$^{12}$C, there is no reason why equivalent states at low density should not 
also exist in heavier $n\alpha$-nuclei, like $^{16}$O, $^{20}$Ne, etc. In a 
mean field picture, i.e. all $\alpha$'s being ideal bosons ( in this context 
remember that the first excited state of an $\alpha$-particle is at 
$\sim$ 20 MeV, by factors higher than in all other nuclei), all $\alpha$'s 
will 
occupy the lowest 0S-state, i.e. 
they will condense into this state. This forms, of course, not a macroscopic 
condensate but 
it can be understood in the same sense as we know that nuclei are superfluid 
because of the presence of a finite number of Cooper pairs. These $\alpha$ 
condensate states are generally close to the $\alpha$ disintegration threshold 
and, therefore, at higher and higher excitation energy as the number 
of $\alpha$'s increases. One may think that, as a consequence, these 
states decay very fast 
but due to their unusual structure, they couple little to ordinary excited 
states and will, thus, have an unusual long life time. On the other hand, 
for example during the cooling process of compact stars \cite{ShapT}, where 
one predicts 
the presence of $\alpha$-particles \cite{ST}, a real macroscopic phase of 
condensed $\alpha$'s may be formed. In the present contribution we will mainly 
concentrate on nuclear systems but we also can think about the possibility of 
quartetting in other Fermi-systems, as mentioned above. One should, however, 
keep in mind that a 
pre-requisite for its existence is, as in nuclear physics, that there exists a 
bound quartet in free space. This is facilitated with the existence of four 
different types of fermions, all attracting one another. For example to form 
quartets with cold atoms one 
could try to trap fermions in four different magnetic substates, a task which 
eventually seems possible \cite{Salo}.
In the next section we will outline the general theory for quartet condensation  and in section 3 we investigate how the binding energy of various 
nuclear clusters changes with density as well as  the critical 
temperature of $\alpha$-particle condensation in infinite matter via an in-
medium four-nucleon equation (Thouless criterion). In section 4, we give some results concerning $\alpha$ condensation in infinite nuclear matter, and in section 5 we treat 
$\alpha$-particle condensation in finite nuclei.  Finally in section 6  we 
conclude with an outlook and further discussions.

\section{General Theory for Quartetting}

The aim will be to develope a theory for quartet condensation which in many 
aspects is analogous to the BCS approach for the condensation of pairs. It 
is, however, evident that a microscopic theory for quartet condensation, 
involving a highly correlated four fermion cluster, is at least by an order 
of magnitude more complicated than the pairing case. It also will turn out 
that the physics is quite different, that is, we will find that quartet 
condensation only exists in the BEC phase and no quartet condensation in weak 
coupling with a very long coherence length, the analogue to the BCS phase of 
pairs, is possible. We will dwell on this aspect in quite some detail. 
In order to set the frame of our approach, let us start to repeat
the BCS approach for pairing. It is well known that the BCS ground state 
wave function can be written as a coherent state \cite{RS}

\begin{equation}
\label{BCSgs}
|\mbox{BCS} \rangle = e^{\sum_{k<k'}z_{kk'}c^+_kc^+_{k'}}|\mbox{vac} \rangle
\end{equation}

\noindent
with $c^+,c$ fermion creation and annihilation operators and indices $k$ 
implying momentum vector, spin, and, eventually other quantum numbers. 
Standard singlet pairing considers $k'=\bar k$, i.e. the pair is at rest and 
the two fermions occupy time reversed states $k$ and $\bar k$. The BCS ground 
state is the vacuum to the quasiparticle destructors 
$\beta_k=u_kc_k - v_kc^+_{\bar k}$ (with $u_k^2 + v^2_k=1$), that is

\begin{equation}
\label{qp-kill}
\beta_k|\mbox{BCS}\rangle = 0
\end{equation}

\noindent
where the standard relation $z_{k\bar k}=v_k/u_k$ is to be used. The 
BCS equations can then be 
obtained by minimising the following mean single particle energy (which can be 
identified with an energy weighted sum rule \cite{RS})

\begin{equation}
\label{sp-average}
e_k=\frac{ \langle \mbox{BCS}|\{ \beta_k,[H-\mu \hat N,\beta^+_k]\} \mbox{BCS}\rangle}{\langle \mbox{BCS}|\{ \beta_k,\beta^+_k\}|\mbox{BCS}\rangle}
\end{equation}

\noindent
where $\{.,.\}$ and $[.,.]$ are anticommutators and commutators, respectively and $\hat N$ and $\mu$ are particle number operator and chemical potential. 
The Hamiltonian is given by

\begin{equation}
\label{hamiltonian}
H=\sum \varepsilon^0_kc^+_kc_k + \frac{1}{4}\sum_{k_1k_2k_3k_4}\bar v_{k_1k_2k_3k_4} c^+_{k_1}c^+_{k_2}c_{k_4}c_{k_3}
 \end{equation}

\noindent
with $\varepsilon^0_k = k^2/(2m)$ the kinetic energy and $\bar v_{k_1k_2k_3k_4}$ the 
antisymmetrised matrix element of the two body interaction. 
Since the quasiparticle operators $\beta^+,\beta$ are obtained from a 
canonical transformation among the $c^+,c$ operators, the transformation can 
be inverted and the $c^+,c$ expressed in terms of $\beta^+,\beta$. Using the 
killing condition, all expectation values contained in the variational 
equations $\delta e_k = 0$ can 
then be expressed in terms of the $u,v$ amplitudes leading directly to the 
usual coupled BCS equations for $u,v$. For later reasons of comparison with the quartet case, let us give the BCS 
in a particular form eliminating the $v$-amplitudes

\begin{equation}
\label{sp-M}
\xi_ku_k + \frac{|\Delta_k|^2}{E_k+\xi_k}u_k = E_ku_k
\end{equation}

\noindent
where $E_k=\sqrt{\xi^2_k+\Delta^2_k}$ is the quasi-particle energy, $\Delta_k=\sum_{k'}V_{kk'}u_{k'}v_{k'}$ the usual 
gap function with $V_{kk'}$ the angle averaged two body force in momentum space, and
$\xi_k = \varepsilon_k - \mu$ with

\begin{equation}
\label{e-HF}
\varepsilon_k = \varepsilon^0_k + \sum_{k'}\bar v_{kk'kk'}n_{k'}
\end{equation}

\noindent
being the Hartree-Fock (HF) single particle energies where $n_k=v^2_k=1-u^2_k$ are 
the single particle occupation numbers. We also want 
to give the equation for the pairing order parameter $\kappa_k \equiv \langle c_{\bar k}c_k\rangle = u_kv_k$ which 
is equivalent to the standard gap equation

\begin{equation}
\label{kappa}
2\varepsilon_k\kappa_k -(1-2n_k)\sum_{k'}V_{kk'}\kappa_{k'}=2\mu\kappa_k
\end{equation}

\noindent
It is seen that (\ref{kappa}) looks similar to a two-body Schr\"odinger 
equation in momentum space, with eigenvalue $2\mu$ and the interaction 
modified by the Pauli blocking factor $(1-2n_k)$. We will see that in the case of 
quartetting, we will get two coupled equations, analogous to (\ref{sp-M}) and (\ref{kappa}).\\

So far for the BCS approach concerning pairs. Let us try to set up an 
analogous procedure for quartets. Obviously we should write for the wave 
function

\begin{equation}
\label{quartet-gs}
|Z\rangle = e^{\frac{1}{4!}\sum_{k_1k_2k_3k_4}Z_{k_1k_2k_3k_4}
c^+_{k_1}c^+_{k_2}c^+_{k_3}c^+_{k_4}}|\mbox{vac}\rangle
\end{equation}

\noindent
where the quartet amplitudes $Z$ are fully antisymmetric (symmetric) 
with respect to 
an odd (even) permutation of the indices. 
The task will now be to find a killing operator for this quartet condensate 
state. Whereas in the pairing case the partitioning of the pair 
operator into a linear combination of a fermion creator and a fermion 
destructor is unambiguous, in the quartet case there exist two ways to 
partition the quartet operator, that is into a single plus a triple or into 
two doubles. Let us start with the superposition of a single and a triple. As 
a matter of fact it is easy to show that ( in the following, we always will assume that all amplitudes are real)

\begin{equation}
\label{q-killer}
q_{\nu} = u^{\nu}_{k_1}c_{k_1} -\frac{1}{3!}\sum v^{\nu}_{k_2k_3k_4}c^+_{k_1}c^+_{k_2}c^+_{k_3}
\end{equation}

\noindent
kills the quartet state under the condition

\begin{equation}
\label{Z=v/u}
Z_{k_1k_2k_3k_4} = \sum_{\nu}(u^{-1})^{\nu}_{k_1}v^{\nu}_{k_2k_3k_4}
\end{equation}

\noindent
However, so far, we barely have gained anything, since above quartet destructor
contains a non-linear fermion transformation which, a priory, cannot be 
handled. Therefore, let us try with a superposition of two fermion pair 
operators which is, in a way, the natural extension of the Bogoliubov 
transformation in the pairing case, i.e. with $Q = \sum [XP - YP^+]$ 
where $P^+ = 
c^+c^+$ is a fermion pair creator. We will, however,  find out that such an 
operator cannot kill the quartet state of Eq.~(\ref{quartet-gs}). 
In analogy to the so-called Self-Consistent RPA (SCRPA) approach 
\cite{Jemai}, we will 
introduce a slightly more general operator, that is

\begin{equation}
\label{pair-bogo}
Q_{\nu} = \sum_{k<k'}[X^{\nu}_{kk'}c_kc_{k'} - Y^{\nu}_{kk'}c^+_{k'}c^+_k]
-\sum_{k_1<k_2<k_3k_4}\eta^{\nu}_{k_1k_2k_3k_4}c^+_{k_1}c^+_{k_2}c^+_{k_3}c_{k_4}
\end{equation}

\noindent
with $X,Y$ antisymmetric in $k,k'$.
Applying this operator on our quartet state, we find $Q_{\nu}|Z\rangle = 0$ 
where the relations between the various amplitudes turn out to be

\begin{equation}
\label{XYZ-relations}
\sum_{k<k'}X^{\nu}_{kk'}Z_{kk'll'} = Y^{\nu}_{ll'}~~~~~~\mbox{and}
~~~~~~~~\eta^{\nu}_{l_2l_3l_4;k'} = \sum_kX^{\nu}_{kk'}Z_{kl_2l_3l_4}
\end{equation}

\noindent
These relations are quite analogous to the ones which hold in the case of the 
SCRPA approach \cite{Jemai}. One also notices that the 
relation between $X, Y, Z$ amplitudes is similar in structure to the 
one of BCS theory for
pairing. As with SCRPA, in order to proceed, we have to approximate the 
additional
$\eta$-term. The quite suggestive recipe is to replace in the $\eta$-term of Eq.~(\ref{pair-bogo}) the density operator $c^+_{k'}c_k$ by 
its mean value $\langle Z|c^+_{k'}c_k|Z\rangle/\langle Z|Z\rangle \equiv 
\langle c^+_{k'}c_k\rangle  =
\delta_{kk'}n_k$, i.e.  $c^+_{k_1}c^+_{k_2}c^+_{k_3}c_{k_4} \rightarrow c^+_{k_1}c^+_{k_2}n_{k_3}\delta_{k_3k_4}$  where we supposed that we work in the basis where the single 
particle density matrix is diagonal, that is, it is given by the occupation 
probabilities $n_k$. This approximation, of course, violates the Pauli 
principle but, 
as it was found in applications of SCRPA \cite{Jemai}, we suppose that 
also here 
this violation will be quite mild (of the order of a couple of percent). With 
this approximation, we see that the $\eta$-term only renormalises the $Y$ 
amplitudes and, thus, the killing operator boils down to a linear super
position of a fermion pair destructor with a pair creator. This can then be 
seen as a Hartree-Fock-Bogoliubov (HFB) transformation of fermion pair 
operators, i.e., pairing of 'pairs'. Replacing the pair operators by 
ideal bosons as done in RPA, would 
lead to a standard bosonic HFB approach \cite{BR} (see also, \cite{RS},ch.9). 
Here, however, we will stay with the 
fermionic description and elaborate an HFB theory for fermion pairs. For this, 
we will suppose that we can use the killing property $Q_{\nu}|Z\rangle = 0$ 
even with the approximate $Q$-operator. As with our experience from SCRPA, 
we assume that this violation of consistency is weak.\\

Let us continue with elaborating our just defined frame. We will then use for the 
pair-killing operator 

\begin{equation}
\label{approxi-Q}
Q_{\nu} = \sum_{k<k'}[X^{\nu}_{kk'}c_kc_{k'} 
- Y^{\nu}_{kk'}c^+_{k'}c^+_k]/N^{1/2}_{kk'} 
\end{equation}

\noindent
with (the approximate) property $Q|Z\rangle=0$ and the first 
relation in (\ref{XYZ-relations}). The normalisation factor $N_{kk'}=
|1-n_k-n_{k'}|$ has been introduced so that $<[Q,Q^+]>= 
\frac{1}{2}\sum (X^2-Y^2) =1$, 
i.e., the quasi-pair state $Q^+|Z\rangle$ and the $X, Y$ amplitudes 
being normalised to one. In analogy with (\ref{sp-average}), we now 
will minimise the following energy weighted sum rule 

\begin{equation}
\label{pair-sum-rule}
\Omega_{\nu} = \frac{\langle Z| [Q_{\nu},[H - 2\mu \hat N,Q^+_{\nu}]]|Z\rangle}
{\langle Z| [Q_{\nu},Q^+_{\nu}]|Z\rangle}
\end{equation}

\noindent
The minimisation with respect to $X, Y$ amplitudes leads to

\begin{equation}
\label{f-pair-gap}
\begin{pmatrix}{\mathcal {\bf H}}&{\bf \Delta}^{(22)}\\-{{\bf \Delta}^{(22)}}^+&
-{\mathcal {\bf H}}^*
\end{pmatrix}
\begin{pmatrix}X^{\nu}\\Y^{\nu} \end{pmatrix}
=\Omega_{\nu}\begin{pmatrix}X^{\nu}\\Y^{\nu} \end{pmatrix}
\end{equation}

\noindent
with (we eventually will consider a symmetrized double commutator in ${\bf H}$)

\begin{eqnarray}
\label{A-matrix}
 {\mathcal {\bf H}}_{k_1k_2,k'_1k'_2} &=& \langle [c_{k_2}c_{k_1},[H-2\mu \hat N,c^+_{k'_1}c^+_{k'_2}]]\rangle/(N^{1/2}_{k_1k_2}N^{1/2}_{k'_1k'_2})\nonumber\\
&=& (\xi_{k_1} + \xi_{k_2})\delta{k_1k_2,k'_1k'_2} +N^{-1/2}_{k_1k_2}N^{-1/2}_{k'_1k'_2}  \{N_{k_1k_2}\bar v_{k_1k_2k'_1k'_2}N_{k'_1k'_2}\nonumber\\ 
&+&  [ (\frac{1}{2}\delta_{k_1k'_1}\bar v_{l_1k_2l_3l_4}C_{l_3l_4k'_2l_1}+\bar v_{l_1k_2l_4k'_2}C_{l_4k_1l_1k'_1} )-(k_1 \leftrightarrow k_2) ] - [k'_1 \leftrightarrow k'_2] \}
\end{eqnarray}

\noindent
where 

\begin{equation}
\label{corr-fct}
C_{k_1k_2k'_1k'_2}=\langle c^+_{k'_1}c^+_{k'_2}c_{k_2}c_{k_1}\rangle -n_{k_1}n_{k_2}\delta_{k_1k_2k'_1k'_2}
\end{equation}

\noindent
is the two body correlation function and

\begin{eqnarray}
\label{B-matrix}
{\bf \Delta}^{(22)}_{k_1k_2,k'_1k'_2} &=& - \langle [c_{k_2}c_{k_1},[H-2\mu \hat N ,c_{k'_1}c_{k'_2}]]\rangle/(N^{1/2}_{k_1k_2}N^{1/2}_{k'_1k'_2})\nonumber\\
&=& N^{-1/2}_{k_1k_2}[( \Delta_{k_1k'_2;k'_1k_2} - k_1 \leftrightarrow k_2 ) - (k'_1 \leftrightarrow k'_2) ]N^{-1/2}_{k'_1k'_2}
\end{eqnarray}

\noindent
with

\begin{equation}
\label{a-gap}
 \Delta_{k_1k'_2;k'_1k_2}=
\sum_{l<l'}\bar v_{k_1k'_2ll'}\langle c_{k'_1}c_{k_2}c_{l'}c_l\rangle 
\end{equation}

\noindent
In (\ref{f-pair-gap}) the matrix multiplication is to be understood  
as $\sum_{k'_1<k'_2}$ 
for restricted summation (or as 
$\frac{1}{2}\sum_{k'_1k'_2}$ for unrestricted summation ) . We see 
from (\ref{B-matrix}) and (\ref{a-gap}) that the bosonic gap 
${\bf \Delta}^{(22)}$ involves the quartet order parameter quite in 
analogy to the 
usual gap field in the BCS case. The ${\bf H}$ operator in (\ref{f-pair-gap}) 
has already been discussed in \cite{NPA628} in connection with SCRPA in 
the particle-particle channel. Equation (\ref{f-pair-gap}) has the typical 
structure of a bosonic HFB equation but, here, for fermion pairs, instead of 
bosons. It remains the task to close those HFB equations in expressing all 
expectation values involved in the ${\bf H}$ and ${\bf \Delta}^{(22)}$ fields 
by the 
$X, Y$ amplitudes. This goes in the following way. Because of the HFB 
structure of (\ref{f-pair-gap}), the $X, Y$ amplitudes obey the usual 
orthonormality relations, see \cite{RS}. Therefore, one can invert 
relation (\ref{approxi-Q}) 
to obtain

\begin{equation}
\label{inversion}
c^+_{k'}c^+_k = N^{1/2}_{kk'}\sum_{\nu}[X^{\nu}_{kk'}Q^+_{\nu} + Y^{\nu}Q_{\nu}]~~~~~~~~~~~(k<k')
\end{equation}

\noindent
and by conjugation the expression for $cc$.
With this relation, we can calculate all two body correlation functions 
in (\ref{B-matrix}) and (\ref{A-matrix}) in terms of $X, Y$ amplitudes. 
This is achieved in commuting the 
destruction operators $Q$ to the right hand side and use the killing property. 
For example, the quartet order parameter in the gap-field (\ref{a-gap}) is 
obtained as $\langle c_{k'_1}c_{k_2}c_{l'}c_l\rangle = 
N^{1/2}_{k'_1k_2}\sum_{\nu}X^{\nu}_{k_2k'_1}Y^{\nu}_{ll'}N^{1/2}_{ll'}$. 
Remains the task how to link the occupation numbers 
$n_k = \langle c^+_kc_k\rangle$ to the $X, Y$ amplitudes. Of course, that is 
where our partitioning of the quartet operator into singles and triples comes 
into play. Therefore, let us try to work with the operator (\ref{q-killer}). 
First, as a side-remark, let us notice that if in (\ref{q-killer}) we replace 
$c^+_{k_1}c^+_{k_2}$ by its expectation value which is the pairing tensor, we are 
back to the standard Bogoliubov transformation for pairing. Here we want to 
consider quartetting and, thus, we have to keep the triple operator fully. 
Minimising, as in (\ref{sp-average}) an average single particle energy, we 
arrive at the following equation for the amplitudes $u, v$ in (\ref{q-killer})

\begin{equation}
\label{sp-eq}
\begin{pmatrix}\xi&{\bf \Delta}^{(13)}\\{{\bf \Delta}^{(13)}}^+&-{\mathcal N}{\mathcal H}^*
\end{pmatrix}
\begin{pmatrix}u\\v \end{pmatrix}
=E\begin{pmatrix}1&0\\0&{\mathcal N} \end{pmatrix}
\begin{pmatrix}u\\v \end{pmatrix}
\end{equation}

\noindent
with (we disregard pairing, i.e., $\langle cc\rangle$ amplitudes)

\begin{equation}
\label{13-gap}
{\bf \Delta}^{(13)}_{k;k_1k_2k_3}= \Delta_{kk_3;k_2k_1} -
[(k_2 \leftrightarrow k_3) - (k_1 \leftrightarrow k_2)]
\end{equation}

\noindent
and

\begin{equation}
\label{3body-H}
({\mathcal N}{\mathcal H}^*)_{k_1k_2k_3;k'_1k'_2k'_3}=\langle \{ c^+_{k_3}c^+_{k_2}c^+_{k_1},[H-3\mu\hat N,
c_{k'_1}c_{k'_2}c_{k'_3}]\}\rangle
\end{equation}

\begin{equation}
\label{3body-N}
{\mathcal N}_{k_1k_2k_3;k'_1k'_2k'_3}=\langle \{ c^+_{k_3}c^+_{k_2}c^+_{k_1},
c_{k'_1}c_{k'_2}c_{k'_3}\}\rangle 
\end{equation}

\noindent
 We will not give ${\mathcal H}$ in full because it is a very complicated 
expression involving self-consistent determination of three-body densities. 
To lowest order in the interaction it is given by

\begin{equation}
\label{H33}
{\mathcal H}_{k_1k_2k_3;k'_1k'_2k'_3}= (\xi_{k_1} +\xi_{k_2}+\xi_{k_3})\delta_{k_1k_2k_3,k'_1k'_2k'_3} + 
[(1 - n_{k_1}-n_{k_2})\bar v_{k_1k_2k'_1k'_2}\delta_{k_3k'_3} + ~~\mbox{permutations}]
\end{equation}

\noindent
where $\delta_{k_1k_2k_3,k'_1k'_2k'_3}$ is the fully antisymmetrised three fermion Kronecker symbol.
Even this operator is still rather complicated for numerical applications and 
mostly one will replace the correlated occupation numbers by their free Fermi-
Dirac steps, i.e., $n_k \rightarrow n^{0}_k$. To this order the three body 
norm in (\ref{3body-N}) is given by

\begin{equation}
{\mathcal N}_{k_1k_2k_3;k'_1k'_2k'_3} \simeq [\bar n^0_{k_1}\bar n^0_{k_2}\bar n^0_{k_3}
+n^0_{k_1} n^0_{k_2} n^0_{k_3}]\delta_{k_1k_2k_3,k'_1k'_2k'_3}
\end{equation}

\noindent
with $\bar n^0 = 1-n^0$.
In principle this effective 
three-body Hamiltonian leads to three-body bound and scattering states. In our 
application to nuclear matter given below, we will make an even more drastic 
approximation and completely neglect the interaction term in the three-body 
Hamiltonian. Eliminating under this condition the $v$-amplitudes from 
(\ref{sp-eq}), one can write down the following effective single particle 
equation

\begin{equation}
\label{eff-sp-field}
\xi_ku_k^{(\nu)} + \sum_{k_1<k_2<k_3k'}\frac{{\bf \Delta}^{(13)}_{k,k_1k_2k_3}(\bar n^{0}_{k_1}\bar n^{0}_{k_2}\bar n^{0}_{k_3} + n^{0}_{k_1}n^{0}_{k_2}n^{0}_{k_3}){\bf \Delta}^{{(13)}^*}_{k_3k_2k_1k'}}{E_{\nu} + \xi_{k_1}+\xi_{k_2}+\xi_{k_3}}u_{k'}^{(\nu)} = E_{\nu}u_k^{(\nu)}
\end{equation}

\noindent
The occupation numbers are given by

\begin{equation}
\label{occ-numbers}
n_k =1-\sum_{\nu}|u^{(\nu)}_k|^2
\end{equation}

\begin{figure}
\begin{center}
\includegraphics[width=7cm]{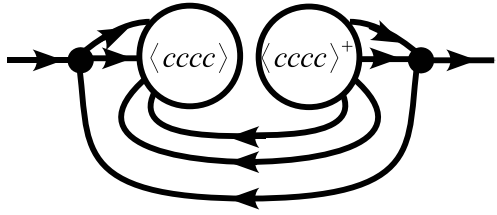}
\caption{\label{fig-dd-mass}Schematic representation of self-energy for 
$\alpha$ particle condensation in an uncorrelated nucleon gas.}
\end{center}
\end{figure}

\noindent
The effective single particle field in (\ref{eff-sp-field}) is grapphically interpreted in Fig.~\ref{fig-dd-mass}.
The gap-fields in (\ref{eff-sp-field}) are then to be calculated as 
in (\ref{13-gap}) and (\ref{a-gap}) 
with (\ref{inversion}) and the system of equations is fully closed. 
This is quite in parallel to the pairing case. 
In cases, 
where the quartet consists out of four different fermions and in addition is 
rather strongly bound, as this will be the case for the $\alpha$ particle in 
nuclear physics, one still can make a very good but drastic simplification: 
one writes the quartic order parameter as a translationally invariant product 
of four times the same single particle wave function in momentum space. We 
will see, how this goes below when we apply our theory to $\alpha$ particle 
condensation in nuclear matter. Comparing the effective single particle field 
in (\ref{eff-sp-field}) with the one of standard pairing, Eq.~(\ref{sp-M}), we 
find strong analogies but also several 
structural differences. The most striking is that in the quartet case Pauli 
factors figure in the numerator of (\ref{eff-sp-field}) whereas this is not the case for pairing. 
In principle in the pairing case, they are also there, but since 
$\bar n_k +n_k = 1$, they drop out. This difference has quite dramatic 
consequences between the pairing and the quartetting case. Namely when the 
chemical potential $\mu$ changes from negative (binding) to positive, the 
implicit three hole level density

\begin{equation}
\label{eq:3h-level}
g_{3h}(\omega)=\sum_{k_1k_2k_3}(\bar n^0_{k_1}\bar n^0_{k_2}\bar n^0_{k_3} + n^0_{k_1}n^0_{k_2}n^0_{k_3})\delta(\omega+\xi_{k_1}+\xi_{k_2}+\xi_{k_3})
\end{equation}

\noindent
passes through zero at $\omega-3\mu=0$ because phase space constraints and 
energy conservation cannot be fullfilled simultaneously at that point. This 
is not at all the case for the single particle level density 

\begin{equation}
\label{1h-level}
g_{1h}(\omega) = \sum_k(\bar n^0_k +n^0_k)\delta(\omega + \xi_k)~~
=~~\sum_k\delta(\omega-\xi_k)
\end{equation}

\noindent
entering implicitly in the pairing case, since, as indicated 
$\bar n^0_k + n^0_k =1$. Therefore, for positive $\mu$, in the case of quartetting 
essentially no correlations around the Fermi energy ($\omega-3\mu =0$) can build up and, 
therefore, there is no quartet condensation for positive chemical potential. 
Quartetting only exists on the BEC side where $\mu$ is negative and, thus, 
$n^0_k=0$ and no structural difference exists between the pairing 
case and the quartetting one for $\mu < 0$. We will show explicit examples 
of level densities in section 4.\\

So far, we have discarded possible existence of ordinary pairing mixed with quartetting. 
Formally, this can easily be achieved in writing everywhere quasiparticle operators instead of particle ones. 
For example the ground state (\ref{quartet-gs}) is then given by

\begin{equation}
\label{quasi-Z}
|Z\rangle = e^{\frac{1}{4!}\sum Z_{k_1k_2k_3k_4}\beta^+_{k_1}\beta^+_{k_2}\beta^+_{k_3}\beta^+_{k_4}}|\mbox{BCS}\rangle
\end{equation}

\noindent
and the killing operator

\begin{equation}
\label{quasi-kill}
Q_{\nu} = \sum_{k<k'}[X^{\nu}\beta_{k'}\beta_k - Y^{\nu}_{kk'}\beta^+_k\beta^+_{k'}]/N_{kk'} 
\end{equation}

\noindent
and analogously for the superposition of singles with triples. 
Of course, if the amplitudes $Z$ in (\ref{quasi-Z}) are zero, we 
are back to ordinary BCS theory of pairs. All the algebra goes 
through as before, only we have to deal in the equations of motion 
with the Hamiltonian in the quasi-particle representation (see \cite{RS}, 
App. E).
This description of quartetting with quasiparticles shall be worked
out in the future.  It may help to understand the precise nature of 
the transition from quartetting to pairing when $\mu$ changes from 
negative to positive values. It is worth noticing that (\ref{quasi-kill}) 
is exactly the ansatz used to derive quasi-particle RPA (QRPA) \cite{RS} 
what implies a linearised equation of motion. Extending to self-consistent 
QRPA \cite{Jemai} will describe pairing and quartetting consistently 
within the same formalism. \\

Of course, the description of quartetting  can also be obtained from 
an equivalent Gorkov-type or Green's function approach. Since this may
shed a complementary light on the theory, we very briefly want to 
sketch how this goes. Let us intruduce two types of matrix Green's
functions

\begin{equation}
\label{13-GF}
{\mathcal G}^{(13)}=
\begin{pmatrix}G^{(11)}&G^{(13)}\\{G^{(13)}}^+&G^{(33)} \end{pmatrix}
\end{equation}

\noindent
with the usual definition of time ordered Green's functions at zero temperature 
to be found, e.g., in \cite{FW},\cite{RS}

\[ G^{(11)}_{kk'} = -i\langle Tc_k(t)c^+_{k'}(t')\rangle ;~~~~~~~~~~
 G^{(13)}_{k;k'_1k'_2k'_3}=-i \langle Tc_k(t)(c_{k'_1}c_{k'_2}c_{k'_3})_{t'}\rangle \]
\[ G^{(13)}_{k_1k_2k_3;k'}=i\langle T(c^+_{k_3}c^+_{k_2}c^+_{k_1})_tc^+_{k'}(t')\rangle ;~~~~~~~~~~
 G^{(33)}_{k_1k_2k_3;k'_1k'_2k'_3}=i\langle T(c^+_{k_3}c^+_{k_2}c^+_{k_1})_t(c_{k'_1}c_{k'_2}c_{k'_3})_{t'}\rangle \]

\noindent
and

\begin{equation}
\label{22-GF}
{\mathcal G}^{(22)}=
\begin{pmatrix}G^{(22)}_{-+}&G^{(22)}_{--}\\G^{(22)}_{++}&G^{(22)}_{+-}
\end{pmatrix}
\end{equation}

\noindent
with
\[ G^{(22)}_{-+,~k_1k_2;k'_1k'_2}= -i \langle P^{(-)}_{k_1k_2}(t)P^{(+)}_{k'_1k'_2}(t')\rangle \]

\noindent
and $P^{(-)}_{k_1k_2}\equiv c_{k_1}c_{k_2}$, and $P^{(+)} = {P^{(-)}}^+$. The other 
Green's functions in the matrix of (\ref{22-GF}) are defined analogously. With
the equation of motion method for fermion cluster Green's functions, see \cite{NPA628}, one then 
can write down a Dyson equation for these matrix Green's functions

\begin{equation}
\label{matrix-GF}
{\mathcal G} = {\mathcal G}^{(0)} + {\mathcal G}^{(0)}\Sigma^{(0)}{\mathcal G}
\end{equation}

\noindent
where ${\mathcal G}$ is either the matrix Green's function (\ref{13-GF}) 
or (\ref{22-GF}) and $\Sigma^{(0)}$ is a corresponding instantaneous part
of the exact self-energy. ${\mathcal G}^{(0)}$ is the free 
Green's function, as usual. The self-energies $\Sigma^{(0,22)}$ and $\Sigma^{(0,13)}$ 
can easily be read off from the effective 2x2 Hamiltonians in (\ref{f-pair-gap}) and (\ref{sp-eq}).
The one and two body density matrices entering these Hamiltonians are then given
by the residua of the various Green's functions and the system of equations
is again closed. This is completely equivalent to the system of equations we 
established before, using the killing operators. 
The Green's function formalism 
has the advantage of easily being generalizable to finite temperature.\\

This ends the formal aspects of quartetting theory. We will give some further insight when we describe the
applications to nuclear systems in the next sections.

\section{Nuclear clusters in the medium and critical temperature of 
quartetting in nuclear matter}


As mentioned in the preceding sections, we will apply our theory for 
quartetting to nuclear matter and also to finite nuclei. In this context it 
may be useful to study first the behaviour of a single fermion pair, e.g. the 
deuteron, and a single quartet, i.e., the $\alpha$ particle in a gas of 
uncorrelated nucleons. For example the in-medium deuteron equation is given by

\begin{equation}
\label{two_part_bind}
\left[\varepsilon_{k_1}+\varepsilon_{k_2}-E_{d,P}\right] \psi_{d,P}(12) +
\sum_{k'_1<k'_2}[1-f_{k_1}-f_{k_2}]\,\,\bar v_{k_1k_2k'_1k'_2} 
 \psi_{d,P}(1'2')=0.
\end{equation}

\noindent
with $P$ the total momentum, $f_k$ the Fermi-Dirac distribution at finite 
temperature (equivalent to the previously defined Fermi step $n^0_k$ at zero temperature, and $\varepsilon_k$ the HF single particle energies 
defined in (\ref{e-HF}).
This {\it effective wave equation} describes bound states as well as
scattering states. The onset of pair condensation is achieved when the 
binding energy $E_{d,P=0}$ coincides with $2 \mu$.

Similar equations have
been derived from the Green function approach for the case of nucleon numbers $A = 3$ and $A = 4$,
describing triton/helion ($^3$He) nuclei
as well as $\alpha$-particles in nuclear matter. The effective wave
equation contains in mean field approximation the Hartree-Fock
self-energy shift of the single-particle energies as well as the Pauli
blocking of the interaction. We give the effective wave equation for the
$\alpha$ particle, 

\begin{eqnarray}
\label{four_part_bind}
 [\varepsilon_{k_1}+\varepsilon_{k_2}+ \varepsilon_{k_3}+\varepsilon_{k_4}
  -E_{\alpha,P}] 
\psi_{\alpha,P}(1234) &+& 
\sum_{k'_1<k'_2}[1-f_{k_1}-f_{k_2}]\bar v_{k_1k_2k'_1k'_2} 
 \psi_{\alpha,P}(1'2'34)\nonumber\\ 
&+& \mbox{permutations}   = 0.
\label{EWE}
\end{eqnarray}

\noindent
A similar equation is obtained for $A=3$.

\begin{figure}[h]
\hspace{0.4cm}
\begin{minipage}[h]{8.5cm}
\includegraphics[width=8.3cm]{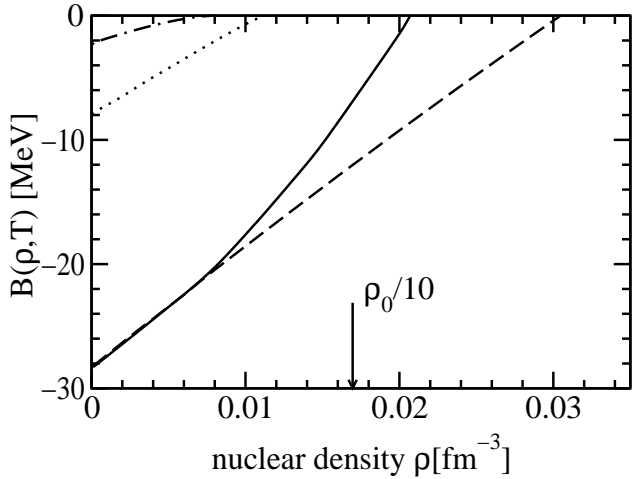}
\end{minipage}
\hfill
\hspace{-0.4cm}
\begin{minipage}[h]{7.cm}
\caption{Shift of binding energies of the light clusters ($d$ - dash dotted, 
$t/h$ - dotted, and $\alpha$ - dashed: perturbation theory, full line : 
non-perturbative Faddeev-Yakubovski equation) in symmetric nuclear
matter as a function of density for given temperature $T = 10$ MeV.}
\label{shifts}
\end{minipage}
\end{figure}

The effective wave equation has been solved using separable potentials
for $A=2$ by integration. For $A=3,4$ we can use a {\it Faddeev
approach} \cite{Beyer}.  The shifts 
of binding energy can also be calculated approximately via perturbation 
theory.  In Fig.~\ref{shifts} we show the shift of the binding 
energy of the light clusters ($d, t/h$ and $\alpha$) in 
symmetric nuclear matter as a function of density for 
temperature $T$ = 10 MeV. 
It is found that the cluster binding energy decreases 
with increasing density.  Finally, at the {\it Mott density} 
$\rho_{A,n,P}^{\rm Mott}(T)$, the bound state is dissolved.  
The clusters are not present at higher densities,  merging into the 
nucleonic medium.  For a given
cluster type characterized by $A,n$, we can also introduce 
the Mott momentum $P^{\rm Mott}_{A,n}(\rho,T)$ in terms of
the ambient temperature $T$ and nucleon density $\rho$,
such that the bound states exist only for 
$P \ge P^{\rm Mott}_{A,n}(\rho,T)$.\\

In general, it is necessary to take account of {\it all bosonic clusters} to
gain a complete picture of the onset of superfluidity.
As is well known, the deuteron is weakly bound compared to
other nuclei.  Higher $A$-clusters can arise that are more stable.  In
this section, we will consider the formation of $\alpha$-particles,
which are of special importance because of their large binding energy
per nucleon ($\sim$7 MeV).  We will not include tritons or helions, which
are fermions and not so tightly bound.  Moreover, we will not consider
nuclei in the iron region, which have even larger binding energy per
nucleon than the $\alpha$-particle and thus constitute, in principle, 
the dominant component at
low temperatures and densities. However, the latter are complex structures of
many particles and are strongly affected by the medium as the density
increases, since their excitation spectrum starts at much lower energy than 
the one of the $\alpha$ particle, for instance. So they are assumed not to 
be of relevance in the density region considered here.
Given the medium-modified 
bound-state energy $E_{4,P}$, the bound-state contribution to 
the EOS is

\begin{equation}
\label{4eos}
\rho_4(\beta,\mu) = \sum_P\left[e^{\beta(E_{4,P} 
- 2 \mu_p-2 \mu_n)} -1\right]^{-1}\,.
\end{equation}

We will not include the contribution of the excited states or
that of scattering states.  Because of the large specific 
binding energy of the $\alpha$ particle, low-density nuclear 
matter is predominantly composed of $\alpha$ particles.
This observation underlies the concept of $\alpha$ matter
and its relevance to diverse nuclear phenomena.
















As exemplified by Eq.~(\ref{EWE}), the 
effect of the medium on the properties of an $\alpha$ particle in
mean-field approximation (i.e., for an uncorrelated medium) is 
produced by the Hartree-Fock self-energy shift and Pauli blocking.
The shift of the $\alpha$-like bound state has been calculated using
perturbation theory as well as by solution of the
Faddeev-Yakubovski equation \cite{Beyer}.  It is found that this
bound state merges with the continuum of scattering states at
a Mott density $\rho_\alpha^{\rm Mott} \approx \rho_0/10$, see 
Fig.~\ref{shifts}. 
The bound states of clusters $d$, $t$,
and $h$ with $A<4$ are already dissolved at the density
$\rho_\alpha^{\rm Mott}$.  Consequently, if we neglect the 
contribution of the four-particle scattering phase shifts 
in the different channels, we can now construct an equation of state 
$\rho(T, \mu) =\rho^{\rm free}(T, \mu) + 
\rho^{{\rm bound}, d}(T, \mu) 
+\rho^{{\rm bound}, \alpha}(T, \mu)$ so that
$\alpha$-particles determine the behavior of symmetric nuclear matter
at densities below $\rho_\alpha^{\rm Mott}$ and temperatures below 
the binding energy per nucleon of the $\alpha$-particle. The 
formation of deuteron clusters alone
gives an incorrect description because 
the deuteron binding energy is small, and the abundance of
$d$-clusters is small compared with that of $\alpha$-clusters. 
In the low density region of the phase diagram, $\alpha$-matter emerges as
an adequate model for describing the nuclear-matter equation
of state.
With increasing density, the medium modifications -- especially Pauli
blocking -- will lead to a deviation of the critical temperature
$T_c(\rho)$ from that of an ideal Bose gas of $\alpha$-particles
(the analogous situation holds for deuteron clusters, i.e., in 
the isospin-singlet channel). At a critical density which more or less 
coincides with the point where the chemical potential turns from negative 
to positive value, the quartet will be quite abruptly dissolved and no 
$\alpha$ particle survives. This is in line with the arguments given above 
concerning the level densities.

 Symmetric nuclear matter is characterized by the equality of
the proton and neutron chemical potentials,
i.e., $\mu_p=\mu_n=\mu$. 
Then an extended Thouless condition based on the relation for the four-body 
in medium wave function, Eq.~(\ref{EWE}), at eigenvalue 4$\mu$
serves to determine the onset of Bose condensation of $\alpha$-like
clusters, noting that existence of a solution of this relation signals
a divergence of the four-particle correlation function. 
An approximate solution has been obtained by a variational
approach, in which the wave function is taken of the following projected 
mean field form \cite{Sogo1}
(see also \cite{RSSN} for another variational ansatz):

\begin{equation}
\label{proj-mf-a}
\psi(1234) = \delta({\bf k}_1 +{\bf k}_2 + {\bf k}_3 + {\bf k}_4 - {\bf K})
\varphi({\bf k}_1)\varphi({\bf k}_2)\varphi({\bf k}_3)\varphi({\bf k}_4)
\chi(1234)
\end{equation}

\noindent
where the $\varphi$'s are single particle wave functions in momentum space to 
be determined variationally out of the in medium four body wave equation 
and $\chi(1234)$ is the scalar spin-isospin function. The delta function is a 
projector on total momentum ${\bf K}$ which, for condensates at rest, is to be taken at ${\bf K}=0$.

\begin{figure}[h]
\hspace{0.7cm}
\begin{minipage}[h]{8.cm}
\includegraphics[width=7.5cm]{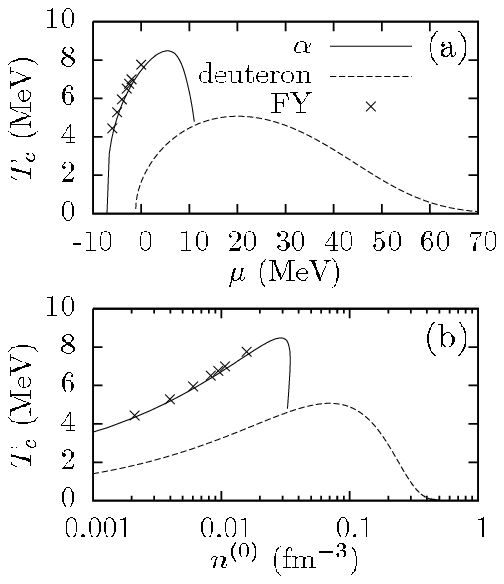}
\end{minipage}
\hfill
\hspace{-0.5cm}
\begin{minipage}[h]{7.5cm}
\includegraphics[width=7.5cm]{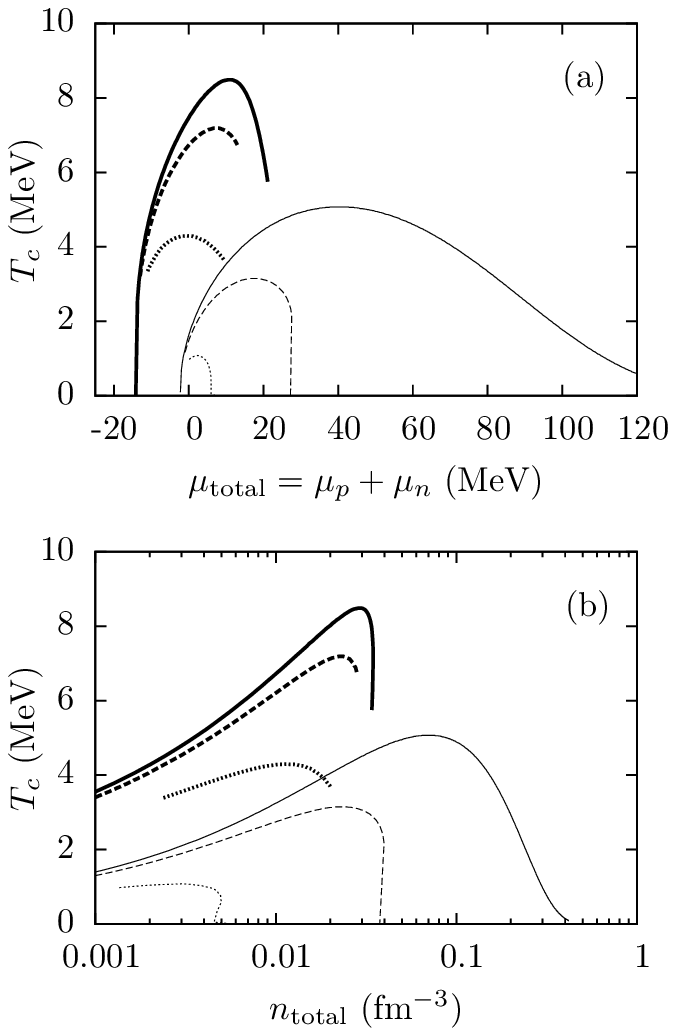}\label{Tc-asym}
\end{minipage}
\caption{(Left) Critical temperatures for $\alpha$ particle and deuteron condensation 
in symmetric nuclear matter as a function of $\mu$ (a) and 
density $n^{(0)}$ (b). (Right) Same as in (Left) but for asymmetric nuclear matter for various 
asymmetry parameters $\delta=(n_n-n_p)/(n_n+n_p)$.  $\delta = 0$ (full line); 
$\delta = 0.5$ (broken line); $\delta = 0.9$ (dotted line).}
\end{figure}

The results for the critical temperature of symmetric and asymmetric 
nuclear matter with a separable force 
which reproduces binding energy and radius 
of a free $\alpha$ particle are presented in Fig.~\ref{Tc-asym}. 
The crosses on the figure indicate an 
exact solution 
of the in medium four body equation with a realistic nucleon-nucleon 
force \cite{Sogo1} \cite{Sogo2}. The excellent agreement with the approximate 
solution justifies 
the choice of our variational wave function. An important feature 
is that at the lowest temperatures, 
Bose-Einstein condensation occurs for $\alpha$ particles rather
than for deuterons.  As the density increases within the low-temperature
regime, the chemical potential $\mu$ first reaches $-7$ MeV, where 
the $\alpha$'s Bose-condense.  By contrast, Bose condensation
of deuterons would not occur until $\mu$ rises to $-1.1$ MeV. In the 
asymmetric case, it is interesting to note that at strong asymmetry $\alpha$ 
particle condensation wins over the deuteron condensation because of the 
strong binding of the $\alpha$'s.

The {\it ``quartetting''} transition temperature sharply 
drops as the rising density approaches the critical Mott 
value ($\mu \simeq 0$) at which the four-body bound states disappear. The deeper reason for this abrupt disappearence is again the same as we discussed above concerning the non-existence of quartet condensation for positive chemical potential, namely the in medium four fermion level density 

\begin{equation}
\label{4-level}
g_4(\omega) = \sum (\bar n^{0}_{k_1}\bar n^{0}_{k_2}\bar n^{0}_{k_3}\bar n^{0}_{k_4}-n^{0}_{k_1}n^{0}_{k_2}n^{0}_{k_3}n^{0}_{k_4})\delta(\omega -\xi_{k_1}-\xi_{k_2}-\xi_{k_3}-\xi_{k_4})
\end{equation}

\noindent
goes, for positive $\mu$ through zero at $\omega -4\mu=0$.
 At that point,
pair formation in the isospin-singlet deuteron-like channel 
comes into play, and a deuteron condensate will exist below the 
critical temperature for BCS pairing up to densities above
the nuclear-matter saturation density $\rho_0$. The reason why the critical temperature for deuterons does not 
break down for positive $\mu$ can again be given via the level density. Namely the two particle in medium level 
density, for pairs {\it at rest}, equals, up to a factor, the one for single particle states.
The critical density at which the $\alpha$ 
condensate disappears is estimated to be around $\rho_0/10 -\rho_0/5$. Therefore, $\alpha$-
particle condensation primarily only exists in the Bose-Einstein-Condensed 
(BEC) phase and there does not seem to exist a phase where the quartets 
acquire a large extension as Cooper pairs do in the weak coupling regime.  
However, 
our variational approach  on which 
this conclusion is based represents only a first attempt at the
description of the transition from quartetting to pairing.  
The detailed nature of this fascinating transition remains 
to be clarified. Further discussions on this phenomenon can be found 
in \cite{Sogo3}. We also should mention that the estimate of our critical temperature for $\alpha$ particle condensation which is based on a generalisation of the Thouless criterion for pairing, is only valid at the upper end of densities for which condensation occurs. For very low densities, the $\alpha$'s should form an ideal Bose gas. For the description of this limit, a theory should be developed which is analogous to the one of Nozi\`eres-Scmitt-Rink \cite{nsr} for pairing. For quartets, this is more involved and has not been worked out so far.


A very intriguing question in relation with non 
existence of an $\alpha$ condensate at higher densities can be asked: is it 
possible that in heavy nuclei an $\alpha$ particle condensate exists in the 
nuclear surface, at least in its fluctuating form? The preformation assumption 
of $\alpha$ particles in the surface to explain $\alpha$ decay may give a 
hint to this.

\section{Alpha-Particle Condensation in Infinite Nuclear Matter at Zero Temperature}

\begin{figure}[h]
\hspace{0.4cm}
\begin{minipage}[h]{10.cm}
\includegraphics[width=7.0cm]{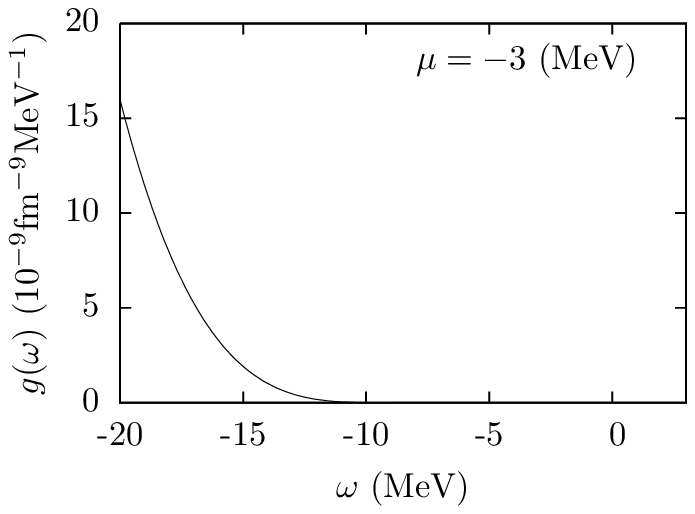}
\includegraphics[width=7.0cm]{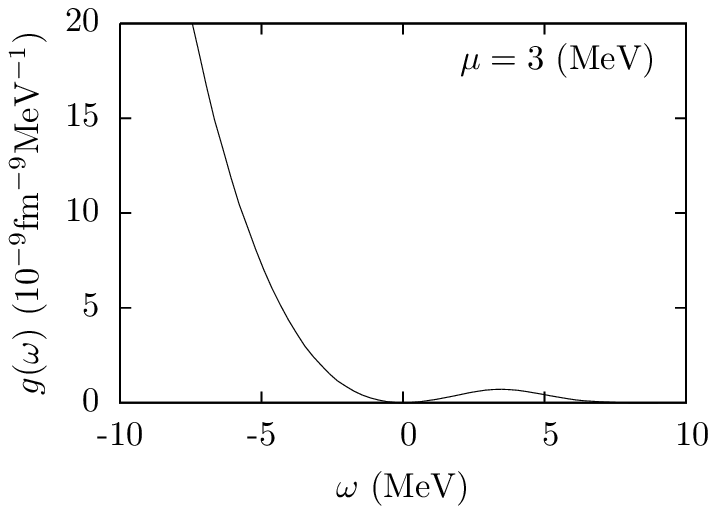}
\end{minipage}
\hfill
\hspace{-0.4cm}
\begin{minipage}[h]{7.cm}
\includegraphics[width=7.0cm]{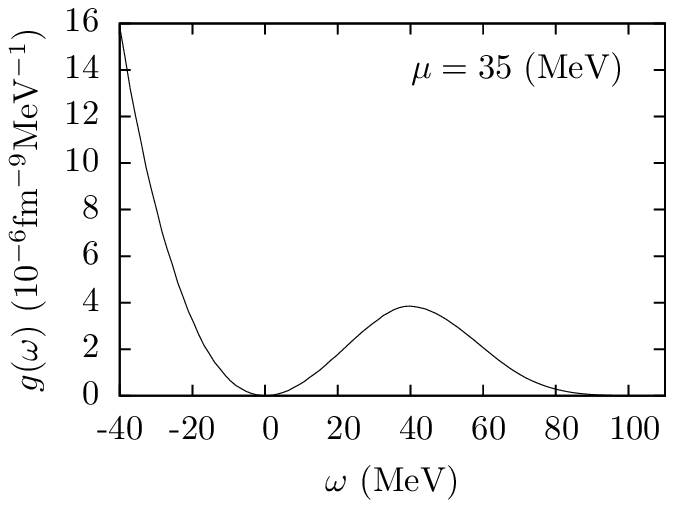}
\caption{\label{fig:3h-level}The 3h level density defined in Eq.~(\ref{eq:3h-level})) for various 
values of the 
chemical potential $\mu$ at zero temperature. The zero on horizontal 
axis is at $3\mu$.}
\end{minipage}
\end{figure}

\begin{figure}
\begin{center}
\includegraphics[width=14.0cm]{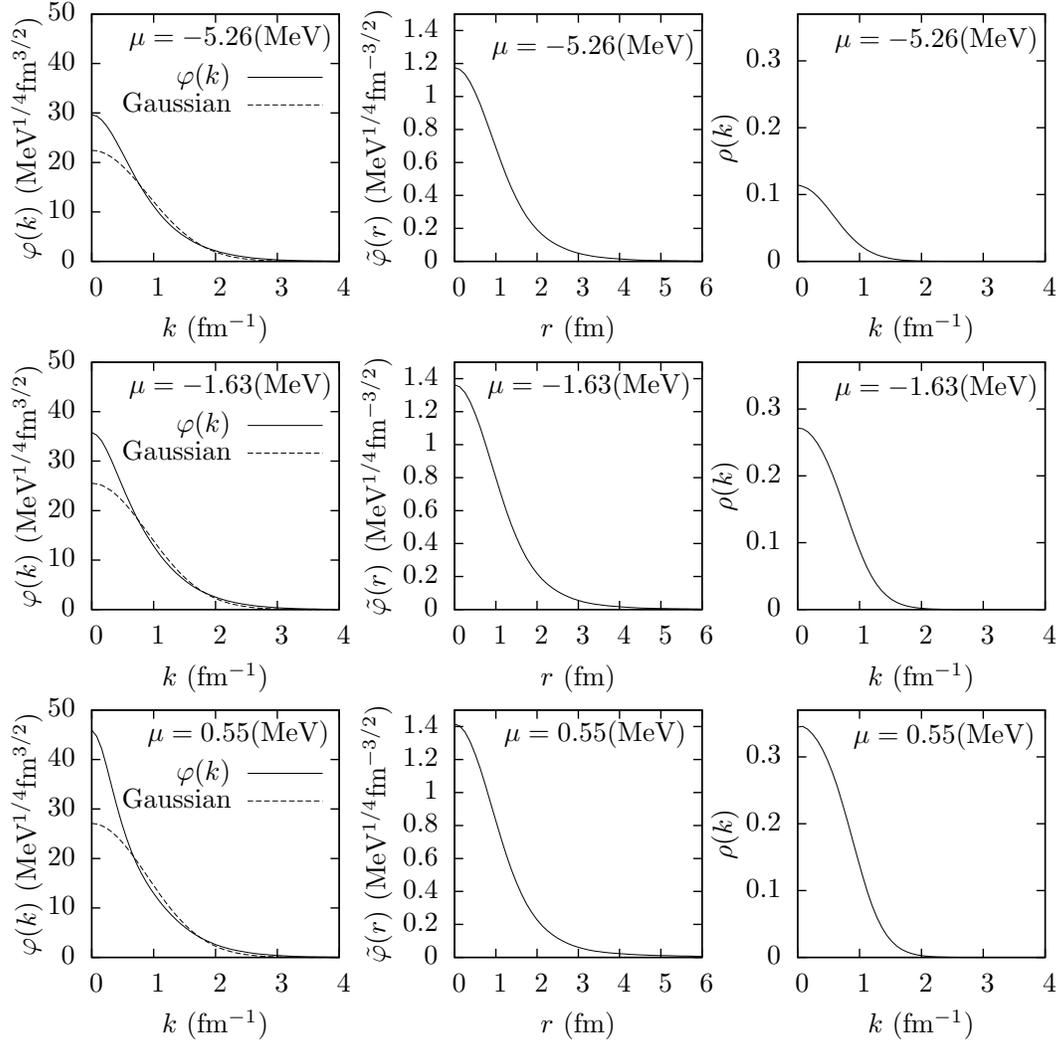}
\caption{\label{fig-phi}
Single particle wave function $\varphi(k)$ for $k$-space (left),
for $r$-space $\tilde \varphi(r)$ (middle), and  
occupation numbers $\rho(k)\equiv n_k$ (right)
at $\mu=-5.26$ (top), $-1.63$ (middle) and $0.55$ (bottom).
The $r$-space wave function $\tilde \varphi(r)$
is derived from the Fourier transform of $\varphi(k)$
by $\tilde \varphi(r)=\int d^3k e^{i\vec k \cdot \vec r}\varphi(k)/(2\pi)^3$.
The dashed line in the left figure correspond to the Gaussian approximation
with same norm and rms momentum as $\varphi(k)$.}
\end{center}
\end{figure}

The equation for the four-body order parameter 
$K_{k_1k_2k_3k_4}\equiv \langle c_{k_1}c_{k_2}c_{k_3}c_{k_4}\rangle$ figuring 
in (\ref{B-matrix}),(\ref{a-gap}),(\ref{13-gap}), obeys, to lowest order 
in the interaction, formally the same equation as the the one which determines 
the critical temperature, namely

\begin{eqnarray}
\label{K4-eq}
(\varepsilon_{k_1}+\varepsilon_{k_2}+\varepsilon_{k_3}+\varepsilon_{k_4})K_{k_1k_2k_3k_4}
&-&\sum_{k'_1<k'_2}[1-n_{k_1}-n_{k_2}]\bar v_{k_1k_2k'_1k'_2}K_{k'_1k'_2k_3k_4}\\ 
&+& \mbox{permutations}=4\mu K_{k_1k_2k_3k_4}
\end{eqnarray}

\noindent
where the occupation numbers $n_k$ should be calculated self-consistently from 
Eq.~(\ref{occ-numbers}). These two coupled equations are analogous to (\ref{kappa}) and (\ref{sp-M}) in the case of pairing. Of course, with an un-approximated 4-body order 
parameter $K$, this would be a tremendously complicated self-consistent 
system of in 
medium 4-body equations to be solved. However, as before, for the critical 
temperature of  $\alpha$ particle condensation, we very effectively 
approximate the order parameter $K$ by our projected mean field ansatz in 
Eq.~(\ref{proj-mf-a}). Therefore the only unknown is now the single particle 
0S wave function $\varphi(k)$. This constitutes a strong 
simplification of the problem. It has been solved in \cite{Sogo3} and 
we refer the 
reader to that reference for further details. We want, however, to show the 
$3h$ level density from that paper, see Fig.~\ref{fig:3h-level}, to demonstrate its vanishing behavior around the Fermi-energy for positive $\mu$. The explicit 
solution of the single particle wave function $\varphi$, 
see Fig.~\ref{fig-phi} . In 
this latter figure, we also give the values of the occupation number 
distribution $n_k$ for various (negative) chemical potentials. It is very 
interesting to see that even for a slightly positive value of $\mu$ 
(i.e., just before the break down of a solution), 
the $n_k$ 
distribution is far from saturation, i.e., from one at $k=0$. 
Its maximum 
value is about 0.35 indicating that one is still far in the BEC side 
when $\alpha$ particle condensation starts to end. \\

\begin{figure}
\begin{center}
\includegraphics[width=12cm]{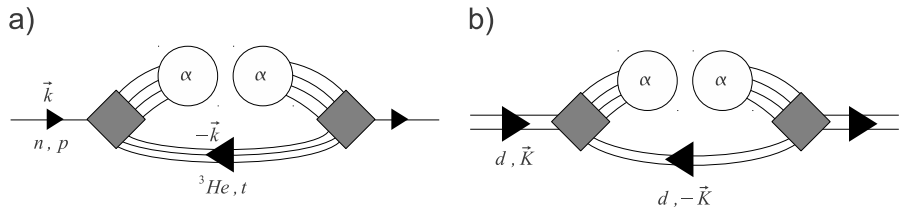}
\caption{\label{fig-dd}Alpha particle creation by scattering of a nucleon on a triton ($^3$He) (left) or of  
two deuterons (right).}\end{center}\end{figure}

Before we will come to the description of $\alpha$ particle condensation in finite nuclei,
let us make some further remarks concerning the approach we have worked out so far. 
This concerns for instance the approximate treatment of the single particle self energy
leading to quartet condensation shown in Fig.~\ref{fig-dd-mass} . This 
process implies that 
$\alpha$ particles are directly formed out of an uncorrelated nucleon gas. 
That the scattering between those particles is neglected may not be the worst
approximation. On the other hand, at least in nuclear physics, there exist 
three nucleon bound states like the triton or $^3$He. In a gas 
of nucleons, they will be 
present and then a scattering process of a nucleon on one of those 'trions' may lead to an
$\alpha$ particle (or the other way round). In addition, there may be deuterons around and two deuterons 
may again form an $\alpha$ particle. All these processes, depicted 
in Fig.~\ref{fig-dd}, are neglected
in our present application and should be included in future studies. One will then have a coherent description of a hot mixture of nucleons, Cooper pairs, deuterons, trions, and $\alpha$-particles which awaits future solution.
Another feature which distinguishes the quartet case from pairing is that, as seen in Fig.~\ref{fig-dd-mass}, there are three hole lines and not just one as for pairing. Because one has to sum over the relative momenta of those three holes, the effective single particle field acquires an imaginary part and no sharp quasi-particle pole developes. More on this has been worked out in \cite{Sogo3}.

\section{Alpha-Particle Condensate States in Self-Conjugate 4n Nuclei}

\begin{figure}[h]
\hspace{0.7cm}
\begin{minipage}[h]{8.cm}
\includegraphics[width=7.5cm]{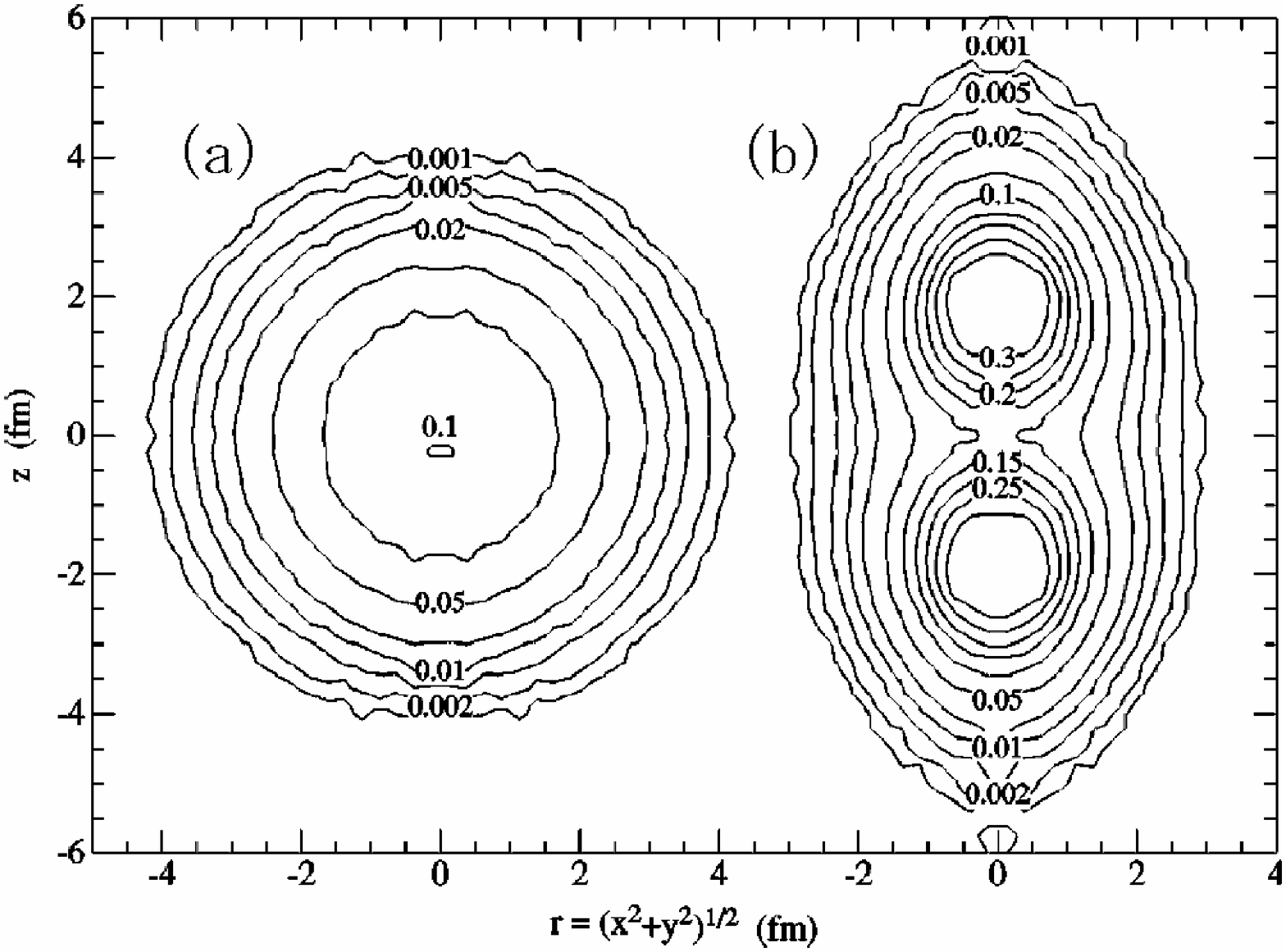}
\end{minipage}
\hfill
\hspace{-0.5cm}

\begin{minipage}[h]{7.5cm}
\caption{Contours of constant density (taken from Ref.~\cite{qmc}),
 plotted in cylindrical coordinates, for $^8$Be$(0^+)$. The left
 side (a) is in the `laboratory'' frame while the right side (b) is in
 the intrinsic frame.}
\label{profiles}
\end{minipage}
\end{figure}

Let us discuss the possibility of quartetting in nuclei. The 
only nucleus having a pronounced $\alpha$-cluster structure 
in its ground state is $^8$Be.  In Fig.~\ref{profiles}(a), the result of an exact calculation of the density 
distribution of $^8$Be in the laboratory frame is shown.  In 
Fig.~\ref{profiles}(b), for comparison, the result
of the same calculation in the intrinsic, deformed frame is displayed.  
We see a 
pronounced two $\alpha$-cluster structure where the two $\alpha$'s are 
$\sim$ 4 fm apart, giving rise 
to a very low average density $\rho \sim \rho_0/3$
as seen in Fig.~\ref{profiles}(a).  As already discussed in the introduction, 
$^8$Be is a rather unusual and unique
nucleus. One may be intrigued by the question, already raised earlier,  
whether 
loosely bound $\alpha$-
particle configurations may not also exist in heavier n$\alpha$-nuclei, at 
least in excited states, naturally close to the n$\alpha$ disintegration 
threshold. Since $\alpha$-particles are rather inert bosons ( first excited 
state at $\sim$ 20 MeV), they then would all condense in 
the lowest S-wavefunction, very much in the same way as do bosonic atoms in 
magneto-optical traps \cite{String}. This
question and exploring related issues of quartetting
in finite nuclei will consume most of the rest of the present 
outline on $\alpha$ particle condensation.
In fact, we will be able to offer strong arguments that the 
$0_2^+$ state of $^{12}$C at 7.654 MeV is a state of 
$\alpha$-condensate nature.  
First, it should be understood that the $0_2^+$ state in 
$^{12}$C is in fact hadronically unstable (as $^8$Be), being 
situated about 300 keV above the three $\alpha$-break up threshold. 
This state is stabilized only by the Coulomb barrier. 
It has a width of $8.7$ eV and a corresponding 
lifetime of $7.6\times 10^{-17}$ s.  As well known, this state is
of paramount astrophysical (and biological!) importance due to 
its role in the creation of $^{12}$C in stellar nucleosynthesis. 
Its existence was predicted in 1953 by the astrophysicist Fred 
Hoyle \cite{hoyle}; his  prediction was confirmed experimentally 
a few years later by Willy Fowler and coworkers at Caltech \cite{fowler}. 
It is also well known that this {\it Hoyle state}, as it is called, 
is a notoriously difficult state for any nuclear theory to explain.
For example, the most modern no-core shell-model 
calculations predict the $0_2^+$ state in $^{12}$C to lie
at around 17 MeV above the ground state -- more than 
twice the actual value \cite{nocore}.  This fact alone tells 
us that the Hoyle state must have a very unusual structure.  
It is easy to understand that, should it indeed have the proposed 
loosely bound three $\alpha$-particle structure, a shell-model 
type of calculation would have great difficulties in reproducing 
its properties.

An important development bearing on this issue took place some 
thirty years ago.  Two Japanese physicists, M. Kamimura \cite{kamimura} 
and K. Uegaki \cite{uegaki}, along with their collaborators, 
almost simultaneously reproduced the Hoyle state from a microscopic 
theory.  They employed a twelve-nucleon wave function together 
with a Hamiltonian containing an effective nucleon-nucleon 
interaction.  At that time, their work did not attract 
the attention it deserved; the true importance of their 
achievement has been appreciated only recently.  The two groups 
started from practically the same ansatz for the $^{12}$C wave 
function, which has the following three $\alpha$-cluster structure:

\begin{equation}
\label{12-RGM}
\langle \vec{r}_1...\vec{r}_{12}|^{12}{\rm C} \rangle = 
{\cal{A}}\left[\chi({{\vec{R}}},\vec{s})\phi_1\phi_2\phi_3\right]\,. 
\end{equation}

\noindent
In this expression, the operator ${\cal A}$ imposes antisymmetry
in the nucleonic degrees of freedom and $\phi_i$, with $i=1,2,3$ for 
the three $\alpha$'s, is an intrinsic $\alpha$-particle wave function of 
prescribed Gaussian form, 

\begin{equation}
\phi(\vec{r}_1,\vec{r}_2,\vec{r}_3, \vec{r}_4) = 
\exp\left\{-\left[(\vec{r}_1-\vec{r}_2)^2
+(\vec{r}_1-\vec{r}_3)^2 +...\right]/{2b^2}\right\}\,, 
\label{alphawf}
\end{equation}

\noindent
where the size parameter $b$ is adjusted to fit the rms of 
the free $\alpha$-particle, and $\chi({{\vec{R}}},\vec{s})$ 
is a yet-to-be determined three-body wave function for the c.o.m.\
motion of the three $\alpha$'s, their corresponding Jakobi coordinates 
being denoted by ${\vec{R}}$ and $\vec{s}$.  The unknown 
function $\chi$ was determined via calculations
based on the Generator Coordinate Method \cite{uegaki} (GCM) 
and the Resonating Group Method \cite{kamimura} (RGM) calculations 
using the Volkov I and Volkov II effective
nucleon-nucleon forces, which fit $\alpha$-$\alpha$ phase 
shifts. The precise solution of this complicated three body problem,
carried out three decades ago, was truly a pioneering achievement,
with results fulfilling expectations. The position of the Hoyle state, 
as well as other properties including the inelastic form factor and 
transition probability, successfully reproduced the experimental 
data.  Other states of $^{12}$C below and around the energy of 
the Hoyle state were also successfully described.  Moreover, it 
was already recognized that the three $\alpha$'s in the Hoyle state 
form sort of a gas-like state.  In fact, this feature had already 
been noted by H.~Horiuchi \cite{hori} prior to the appearance
of Refs.~\cite{kamimura,uegaki}, based on results from the
orthogonality condition model (OCM) \cite{saitoh}. 
All three Japanese research groups concluded from their studies 
that the linear-chain state of three $\alpha$-particles, 
postulated by Morinaga many years earlier \cite{mori}, 
had to be rejected.

Although the evidence for interpreting the Hoyle state 
in terms of an $\alpha$ gas was stressed in the cited papers
from the late 1970's, two important aspects of the situation 
were missed at that time.  First, because the three $\alpha$'s 
move in identical $S$-wave orbits, one is dealing with an
$\alpha$-condensate state what may, in fact, be a general feature in $n\alpha$ 
nuclei. 
The second and most important point is that the complicated three-body wave 
function $\chi({\vec{R}},{\vec{s}})$ for the c.o.m. motion of the three 
$\alpha$'s 
can be replaced by a structurally and conceptually very 
simple microscopic three-$\alpha$ wave function of the 
condensate type, which has practically 100 percent overlap 
with the previously constructed ones \cite{thsr} \cite{cbec} (see also 
Ref.~\cite{Hackenbroich}). We now 
describe this condensate wave function.

We start by examining the BCS wave function of ordinary fermion
pairing, obtained by projecting the familiar BCS 
ground-state ansatz onto an $N$-particle subspace of Fock space.  
In the position representation, this wave function is

\begin{equation}
\langle \vec{r}_1...\vec{r}_N|{\rm BCS}\rangle = {\cal {A} }
\left[\phi(\vec{r}_1,\vec{r}_2)\phi(\vec{r}_3,\vec{r}_4)...\phi(\vec{r}_{N-1}
\vec{r}_N)\right]\,, \label{eq:1}
\end{equation}

\noindent
where $\phi(\vec{r}_1,\vec{r}_2)$ is the Cooper-pair wave function 
(including spin and isospin), which is to be determined variationally  
through the familiar BCS equations.  The condensate character of the BCS 
ansatz is born out by the fact that within the antisymmetrizer $\cal A$,
one has a product of $N/2$ times the same pair wave function $\phi$, 
with one such function for each distinct pair in the reference
partition of $\{1,2,\ldots,N-1,N\}$.  Formally, it is now a simple 
matter to generalize (\ref{eq:1}) to quartet or $\alpha$-particle 
condensation. 
We write

\begin{equation}
\langle \vec{r}_1,\ldots,\vec{r}_N|\Phi_{n\alpha}\rangle = {\cal {A} }\left[ 
\phi_{\alpha}(\vec{r}_1, \vec{r}_2, \vec{r}_3, \vec{r}_4)\phi_{\alpha}
(\vec{r}_5,\ldots , \vec{r}_8)
\cdots \phi_{\alpha}(\vec{r}_{N-3},\ldots, \vec {r}_N)\right]\,, \label{eq:2}
\end{equation}

\noindent
where $\phi_{\alpha}$ is the wave 
function common to all condensed $\alpha$-particles. It may be considered as the number projected quartet wave function introduced in (\ref{quartet-gs}). Of course, 
finding the variational solution for this function
is, in general, extraordinarily more complicated than finding
the Cooper pair-wave function $\phi$ of Eq.~(\ref{eq:1}). 
Even so, in the present case that the $\alpha$-particle is the
four-body cluster involved, and for applications to 
relatively light nuclei, the complexity of the problem can be reduced 
dramatically.  This possibility stems from the fact that 
an excellent variational ansatz for the intrinsic wave function
of the $\alpha$-particle is provided [as in Eq.~(\ref{alphawf})], 
by a Gaussian form with only the size parameter $b$ to be determined. 
The new aspect in \cite{thsr} was that in addition even the center-of-mass 
motion of 
the system of $\alpha$-particles can be described very well 
by a Gaussian wave function with, this time, a size parameter 
$B \gg b$ to account for the motion over the whole nuclear space. 
This is a strong 
technical simplification and, at the same time, underlines the boson 
condensate character of the wave function.  
We therefore write

\begin{equation}
\phi_\alpha(\vec{r}_1,\vec{r}_2,\vec{r}_3,\vec{r}_4) = 
e^{{\displaystyle{-2}}{\scriptstyle\vec{R}^2}{\displaystyle{/B^2}}}
\phi(\vec{r}_1-\vec{r}_2,\vec{r}_1-\vec{r}_3,\cdots)\,,  \label{eq:3}
\end{equation}

\noindent
where $\vec{R}= (\vec{r}_1+\vec{r}_2+\vec{r}_3+\vec{r}_4)/4$ is the c.o.m.\ 
coordinate of one $\alpha$-particle and $\phi(\vec{r}_1-\vec{r}_2,...)$ is 
the same intrinsic $\alpha$-particle wave function of Gaussian form as 
already used in Refs.~\cite{kamimura,uegaki} and given explicitly
in Eq.~(\ref{alphawf}). This wave function has a close relation to the one 
we used in the infinite matter case. To this end, let us Fourier transform the 
infinite matter ansatz (\ref{proj-mf-a}) into real space ( with a projection onto a finite ${\bf K}$)

\begin{equation}
\label{equiv-infty}
\phi^{\infty}_{\alpha,{\bf K}}(\vec{r}_1,\vec{r}_2,\vec{r}_3,\vec{r}_4)= e^{i{\bf K}{\bf R}} \int d^3R'e^{-i{\bf K}{\bf R}'}
\tilde \varphi({\bf r}_1-{\bf R}')\tilde \varphi({\bf r}_2-{\bf R}')\tilde \varphi({\bf r}_3-{\bf R}')
\tilde \varphi({\bf r}_4-{\bf R}')
\end{equation}

\noindent
where $\tilde \varphi$ is the Fourier transform of $\varphi$. Taking for $\varphi$ a Gaussian, one obtains

\begin{equation}
\label{infty-gauss}
\phi^{\infty}_{\alpha,{\bf K}}(\vec{r}_1,\vec{r}_2,\vec{r}_3,\vec{r}_4)= 
e^{i{\bf K}{\bf R}}
\exp\left\{-\left[(\vec{r}_1-\vec{r}_2)^2
+(\vec{r}_1-\vec{r}_3)^2 +...\right]/{2b^2}\right\}
\end{equation}

\noindent
Comparing with (\ref{eq:3}), we, therefore, see that our variational 
ansatz for the $\alpha$ particle 
condensate wave function is of the same spirit as the one we used already 
in the homogeneous case, only the plane wave c.o.m. wave function has, 
naturally, been replaced by a Gaussian. For a small number of $\alpha$ 
particles, it is, of course, important to work with a wave function with a 
definite number of particles and not with the coherent 
state (\ref{quartet-gs}). However, for a larger number of $\alpha$'s 
where the handling of (\ref{eq:2}) becomes more and more difficult 
because of the explicit antisymmetrisation of all nucleons, it may 
be worth to use (\ref{eq:3}) also in our quartet coherent 
state (\ref{quartet-gs}) as a variational wave function with the two 
parameters $B,b$.

Naturally, in Eq.~(\ref{eq:2}) the center of mass $\vec{X}_{\rm cm}$ of 
the three $\alpha$'s, i.e., of the whole nucleus, should be eliminated; 
this is easily achieved by replacing $\vec{R}$ by $\vec{R}-\vec{X}_{\rm cm}$ in 
each of the $\alpha$ wave functions in Eq.~(\ref{eq:2}). 
The 
$\alpha$-particle condensate wave function specified by Eqs.~(\ref{eq:2}) 
and (\ref{eq:3}), proposed in Ref.~\cite{thsr} and  
called the THSR wave function, now depends on only two parameters, 
$B$ and $b$. The expectation value of an assumed microscopic Hamiltonian
$H$,

\begin{equation}
{\cal {H}}(B,b)=\frac{\langle\Phi_{n\alpha}(B,b)|H|\Phi_{n\alpha}(B,b)\rangle}
{\langle \Phi_{n\alpha}|\Phi_{n\alpha}\rangle}\,, \label{eq:4}
\end{equation}

\noindent
can be evaluated, and the corresponding two-dimensional energy surface 
can be quantized using the two parameters $B$ and $b$ 
as Hill-Wheeler coordinates. \index{Hill-Wheeler method}

Before presenting the results, let us discuss the THSR wave 
function in somewhat more detail.  This innocuous-looking variational
ansatz, namely Eq.~(\ref{eq:2}) together with Eq.~(\ref{eq:3}), 
is actually more subtle than it might at first appear. One should 
realize that two limits are incorporated exactly.  One is obtained by 
choosing $B=b$, for which Eq.~(\ref{eq:2}) reduces to a standard 
Slater determinant with harmonic-oscillator single-nucleon 
wave functions, leaving the oscillator length $b$ as the 
single adjustable parameter.  This holds because the 
right-hand-side of expression (\ref{eq:3}), with $B=b$, becomes 
a product of four identical Gaussians, and the antisymmetrization 
creates all the necessary $P$, $D$, etc.\ harmonic oscillator 
wave functions automatically \cite{thsr}.  On the other hand,
when $B \gg b$, the density of $\alpha$-particles is very low, 
and in the limit $B \rightarrow \infty$, the average distance between 
$\alpha$'s is so large that the antisymmetrisation between 
them can be neglected, i.e., the operator $\cal {A}$ in front of 
Eq.~(\ref{eq:2}) becomes irrelevant and can be removed.
In this limiting case, our wave function then describes an ideal gas of 
independent, condensed $\alpha$-particles -- it is a pure product 
state of $\alpha$'s! An elucidating study on this aspect is given in 
Ref.~\cite{yamada1}.\\

Evidently,  in realistic cases the antisymmetrizer 
$\cal {A}$ cannot be neglected, and evaluation of the expectation 
value (\ref{eq:4}) becomes a nontrivial task. 
The Hamiltonian in Eq.~(\ref{eq:4}) was taken to be the one
used in Ref.~\cite{tohsaki_F1}, which features an effective 
nucleon-nucleon force of the Gogny type, with parameters fitted
to $\alpha$-$\alpha$ scattering phase shifts as available about 
fifteen years ago.  This force also leads to very reasonable 
properties of ordinary nuclear matter.  Our theory is therefore 
free of any adjustable parameters.  The energy landscapes 
${\cal {H}}(B,b)$ for various $n\,\alpha$ nuclei are interesting 
in themselves \cite{tohsaki_nara}, but for the sake of brevity 
they are not shown here.\\

\begin{figure}
\begin{center}
\includegraphics[width=6cm]{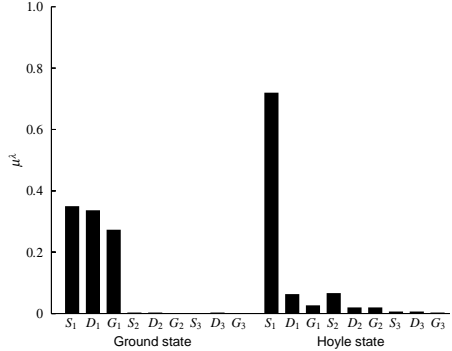}
\caption{\label{3a-occs}$\alpha$ occupation numbers in the ground 
state (left) and Hoyle state (right) of $^{12}$C.}
\end{center}
\end{figure}

As we discussed already, the variational
wave function constructed from the Hill-Wheeler equation based on 
Eqs.~(\ref{eq:2}), (\ref{eq:3}), and (\ref{eq:4}) has practically 
100 percent overlap with the RGM and GCM wave functions constructed in 
Refs.~\cite{kamimura} and \cite{uegaki}, once the same 
Volkov force is used \cite{cbec}. However, one can take fixed optimised values 
for $b$ and $B$ parameters and, then, the corresponding {\it single} THSR wave 
function still has 98 percent squared overlap with the RGM or GCM solutions. 
It is, thus, not astonishing 
that our results are very similar to the RGM and GCM ones.  For $^{12}$C we obtain 
two eigenvalues: the ground state and the Hoyle state. Theoretical values 
for positions, rms values, and transition probabilities are 
given in Table~\ref{tab:1} and compared to the data.  Inspecting
the rms radii, we see that the Hoyle state has a volume 3 to 4 larger 
than that of the ground state of $^{12}$C.  This is the primary 
aspect of the dilute-gas state we highlighted above. Constructing 
a pure-state $\alpha$-particle density matrix $\rho(\vec{R},\vec{R}')$ 
from our wave function, integrating out of the total density matrix 
all intrinsic $\alpha$-particle coordinates, and diagonalizing this 
reduced density matrix, we find that the corresponding $0S$ 
$\alpha$-particle orbit is occupied to 70 percent by the 
three $\alpha$-particles \cite{yamada1,suzuki} whereas the occupation of all 
other states is down by at least a factor of ten, see Fig.~\ref{3a-occs}.  

This is a huge 
percentage, giving vivid support to the view that the Hoyle
state is an almost ideal $\alpha$-particle condensate. 
\index{alpha cluster!condensate}  By way of contrast, we should also mention 
that in the ground state of $^{12}$C, the $\alpha$-particle 
occupation is about equally shared between the $0S$, $0D$, and $0G$ 
orbits, i.e. yielding the shell model limit, clearly invalidating a 
condensate picture of the ground state, see Fig.~\ref{3a-occs} (it is 
important to note that the ground-state energy of $^{12}$C is 
also reasonably reproduced by our theory). We, thus, think that the investigation of the bosonic occupancies is the most adequate way to demonstrate whether a given state can, in nuclei,  be qualified as an $\alpha$ condensate or not.

\begin{table}
\begin{center}
\begin{tabular}{ccccc}
\hline\hline
 &  & condensate w.f. & \raisebox{-1.8ex}[0pt][0pt]{RGM \cite{kamimura}} & \raisebox{-1.8ex}[0pt][0pt]{Exp.} \\
 &  & (Hill-Wheeler) &  &  \\
\hline
\raisebox{-1.8ex}[0pt][0pt]{$E$(MeV)} & $0_1^+$ & $-89.52$ & $-89.4$  & $-92.2$  \\
 & $0_2^+$ & $-81.79$ & $-81.7$  & $-84.6$  \\
\hline
\raisebox{-1.8ex}[0pt][0pt]{$R_{\rm r.m.s.}$(fm)} & $0_1^+$ &   $\ \ \ 2.40$ &   $\ \ \ 2.40$ &   $\ \ \ 2.44$ \\
 & $0_2^+$ &   $\ \ \ 3.83$ &   $\ \ \ 3.47$ &  \\
\hline
$M(0_2^+\rightarrow 0_1^+)$(fm$^2$) &  &   $\ \ \ 6.45$ &   $\ \ \ 6.7$ &   $\ \ \ 5.4$  \\
\hline\hline
\end{tabular}
\caption{Comparison of the binding energies, rms radii $(R_{\rm r.m.s.})$, 
and monopole matrix elements $(M(0_2^+\rightarrow 0_1^+))$ for 
$^{12}$C given by solving Hill-Wheeler equation  
  based on Eq.~(\ref{eq:2}) and by Ref.~\cite{kamimura}. 
  The effective two-nucleon force Volkov No.~2 was adopted 
  in the two cases for which the $3\alpha$ threshold energy is
  calculated to be $-82.04$ MeV.}
\label{tab:1}
\end{center}
\end{table}


\begin{figure}[h]
\hspace{.4cm}
\begin{minipage}[h]{7.cm}
\includegraphics[width=5.cm]{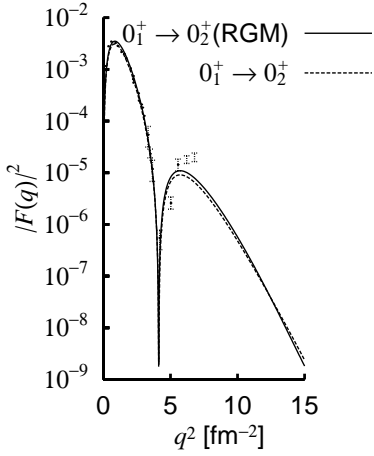}
\end{minipage}
\hfill
\hspace{-0.4cm}
\begin{minipage}[h]{7.cm}
\caption{Experimental values of inelastic form factor in $^{12}$C to
  the Hoyle state are compared with our values and those given by
  Kamimura et al. in Ref.~\cite{kamimura} (RGM).
  In our result, the Hoyle-state wave function is obtained by
  solving the Hill-Wheeler equation based on Eq.~(\ref{eq:2}).
  }
\label{fig:2}
\end{minipage}
\end{figure}

\begin{figure}[h]
\hspace{0.5cm}
\begin{minipage}[h]{7.0cm}
\includegraphics[width=7.0cm]{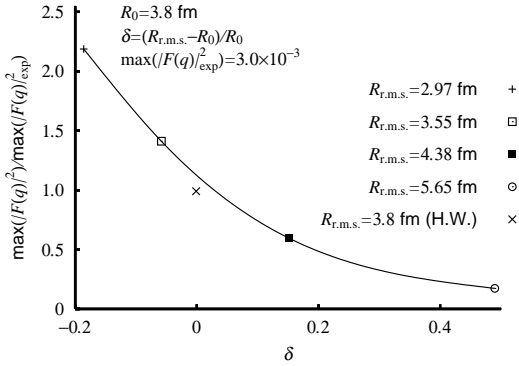}
\end{minipage}
\hfill
\hspace{-0.5cm}
\begin{minipage}[h]{7.0cm}
\caption{The ratio of the value of the maximum height, theory versus
  experiment, of the inelastic form factor,
  i.e. max$|F(q)|^2$/max$|F(q)|_{\rm exp}^2$, is plotted as a
  function of $\delta=(R_{\rm r.m.s.}-R_0)/R_0$.  Here $R_{\rm r.m.s.}$
  and $R_0$ are the rms radii corresponding, respectively, to the
  wave function of Eq.~(\ref{eq:2}) and that obtained by solving
  the Hill-Wheeler equation based on Eq.~(\ref{eq:2}).}
\label{fig:3}
\end{minipage}
\end{figure}

Let us now discuss what to our mind is the most convincing evidence 
that our description of the Hoyle state is the correct one. 
Like the authors of Ref.~\cite{kamimura}, we reproduce very 
accurately the inelastic form factor $0_1^+ \rightarrow 0_2^+$
of $^{12}$C, as shown in Fig.~\ref{fig:2}.  As such, the agreement with 
experiment is already quite impressive in view of the fact that we did not 
use a single adjustable parameter.  Additionally, however, the
following study was made, results from which are presented in 
Fig.~\ref{fig:3}.  We artificially varied the extension of the 
Hoyle state and examined the influence on the form factor.  It 
was found that the overall shape of the form factor shows 
little variation, for example in the position of the minimum. 
On the other hand, we found a strong dependence of
the absolute magnitude of the form factor; Fig.~\ref{fig:3} 
illustrates this behavior with a plot showing the variation of 
the height of the first maximum of the inelastic form factor 
as a function of the percentage change of the rms radius of 
the Hoyle state \cite{cbec}.  It can be seen that a 20 
percent {\it increase of the rms radius} produces a remarkable
decrease of the maximum -- by a factor of 
two! This strong sensitivity of the magnitude of the form 
factor to the size of the Hoyle state enhances our firm 
belief that the agreement with the actual measurement is 
tantamount to a proof that the calculated wide extension 
of the Hoyle state corresponds to reality.
We thus advocate and support the view that
the Hoyle state can be regarded as the ground state of an
$\alpha$-particle condensate \cite{exstB}. We should, however, be aware of the 
fact that it is not an ideal condensate and that the $\alpha$'s occupy the 
lowest S-state only with 70 percent, as already discussed. The major part of 
the correlations comes from the Pauli principle and intermediate $^8$Be 
formation.
It is by the way not clear whether a gas of $\alpha$ particles condenses as 
such or as 'molecules' of $^8$Be. The latter are also bosons, of course.
We furthermore performed a deformed calculation to investigate the 
structure of the $2_2^+$ state in $^{12}$C. The state came at the right energy. 
Our analysis showed that this state essentially corresponds to exciting 
one $\alpha$-particle 
out of the condensate and putting it into the $0D$ orbit. 
Without going into details, we also 
affirm that the width of this state is correctly reproduced 
\cite{fthsr}. It should also be mentioned that this $2_2^+$ state has in our 
calculations an enormous extension with an rms radius of 4.3 fm what 
corresponds approximately to eight times the ground state volume of $^{12}$C 
or also to the size of $^{40}$Ca.   
As a matter of fact the properties of this state have been subject of a vivid 
debate among the experimentalists in the recent past. The situation seems 
clarified now \cite{itoh}\cite{freer0}\cite{Gai}. It is for instance, the 
very nice 
experiment by 
Moshe Gai \cite{Gai} which confirms the properties of the $2_2^+$ state 
beyond any doubt. In that reference it also is given an estimate of the 
radius which agrees with our value. One can talk about an $\alpha$- 
halo state. Further experimental verification of this giant $\alpha$-gas state would be very welcome.\\

\begin{figure}
\includegraphics[width=14cm]{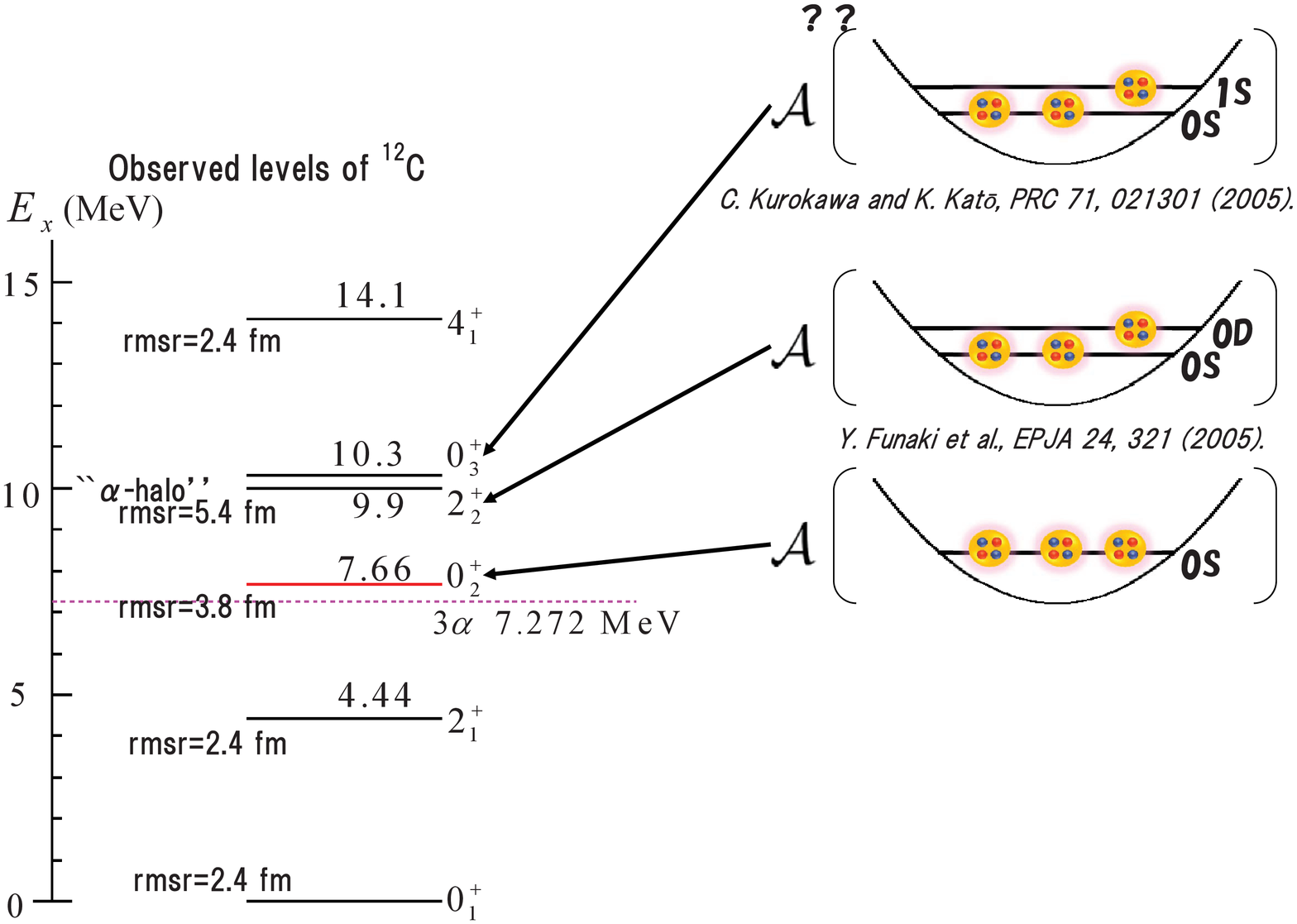}
\caption{\label{12C} Spectrum of $^{12}$C and its interpretation
via an analysis with the THSR wave function concerning 
$\alpha$ cluster states.}
\end{figure}

It is tempting to imagine that the $0_3^+$ 
state which -- experimentally -- is almost degenerate with 
the $2_2^+$ state, is obtained by lifting one $\alpha$-particle 
into the $1S$ orbit.  Initial theoretical studies \cite{kato} 
indicate that this scenario might indeed apply.  However, the
width of the $0_3^+$ state ($\sim$ 3 MeV) is very broad, rendering 
a theoretical treatment rather delicate.  Further investigations 
are necessary to validate or reject this picture.  At any rate, 
it would be quite satisfying if the triplet of states 
(${0_2}^+\,,{2_2}^+\,,{0_3}^+$) could all be explained from the 
$\alpha$-particle perspective, since those three states are 
{\it precisely} the ones which cannot be reproduced within a 
(no core) shell-model approach \cite{nocore}. In Fig.~\ref{12C} we represent
this scenario.

Summarizing our inquiry into the possible role of $\alpha$
clustering in $^{12}$C, we have accumulated enough facts to 
be convinced that the Hoyle state is, indeed, what one may
call to first approximation an $\alpha$-particle condensate state.  At the same 
time, we acknowledge that referring to only three particles 
as a ``condensate'' constitutes a certain abuse of the word.  
However, in this regard it should be remembered that also
in the case of nuclear Cooper pairing, only a few pairs 
are sufficient to obtain clear signatures of superfluidity 
in nuclei!

What about $\alpha$-particle condensation in heavier nuclei?
Once one accepts the idea that the Hoyle state is essentially a 
state of three free $\alpha$-particles held together only by the Coulomb 
barrier, it is hard to see why analogous states would not also exist 
in heavier $n\,\alpha$ nuclei like $^{16}$O, $^{20}$Ne, $^{24}$Mg, etc.
In this respect, it is important to recognize the argument that the 
successively higher excitation energies of $\alpha$ condensate states 
do not necessarily imply very short 
life times because of their very unusual structure, 
having little in common with ordinary nuclear states.
Our calculations on such nuclei systematically 
yield a $0^+$-state close to the $\alpha$-particle disintegration 
threshold.  For example in $^{16}$O we obtain three $0^+$-states \cite{thsr}: 
the ground state at $E_0= -124.8 $ MeV (experimental value: 
127.62 MeV), a second state at excitation energy $E_{{0_2}^+} = 8.8$ MeV, and 
a third one at $E_{{0_3}^+} = 14.1$ MeV. The threshold 
in $^{16}$O is at 14.4 MeV.  Unfortunately, the relevant 
experimental information in $^{16}$O is not nearly so 
complete as in $^{12}$C. In particular, no measurements are
available for transition probabilities of $0^+$-states near 
the threshold or for inelastic form factors.

\begin{figure}[h]
\hspace{0.7cm}
\begin{minipage}[h]{8.cm}
\includegraphics[width=7.5cm]{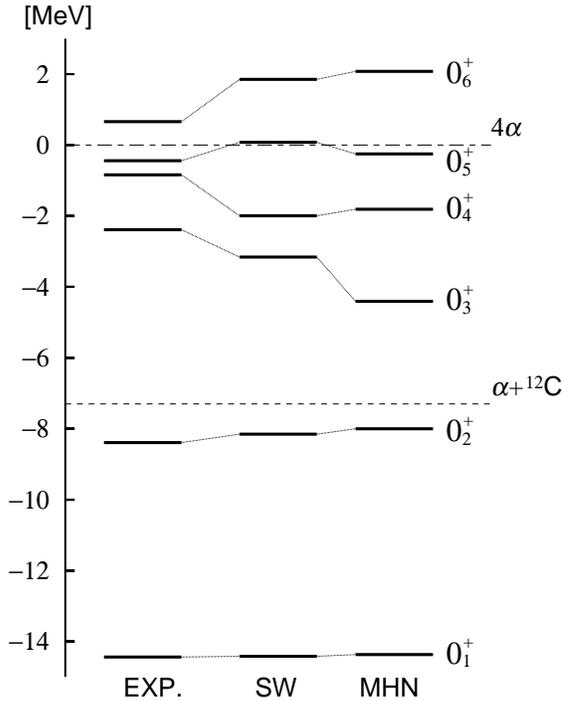}
\end{minipage}
\hfill
\hspace{-0.5cm}
\begin{minipage}[h]{7.5cm}
\caption{Spectrum of first six $0^+$ states in $^{16}$O.
Left: experiment; middle: theory with SW force; right: theory with MHN force; see \cite{ocm16} for more details.}
\label{16O}
\end{minipage}
\end{figure}

In contrast to the situation for $^{12}$C, 
the THSR wave function is certainly not able to describe the 
structure of all $0^+$-states in $^{16}$O lying below the
disintegration threshold.  In $^{12}$C knocking loose one $\alpha$ particle, 
the 
other two are also loosely bound, since what remains is $^{8}$Be. However, 
in $^{16}$O this is not the case. Before reaching a four $\alpha$ particle 
gas state, there will appear configurations where one $\alpha$ particle orbits 
around a $^{12}$C core in its ground state or in excited states of particle-
hole type. A case in point is the first excited state 
in $^{16}$O, i.e., the ${0_2}^+$-state at 6.06 MeV, which is 
believed to have a structure corresponding to an $\alpha$-particle 
orbiting in an $S$ wave around a $^{12}$C core in its ground state. 
Such a configuration is clearly missing from our wave function 
(\ref{eq:2}).  As a matter of fact a calculation for the first six $0^+$ 
states has been performed in the meanwhile employing a somewhat different, 
not completely microscopic approach \cite{ocm16}. This method is the so-called 
'orthogonality condition model (OCM)' which generally works also quite well
for the description of cluster states. 
The 
novelty with respect to former calculations of $^{16}$O states with this method 
was 
that the configuration space was strongly enlarged. In  Fig.~\ref{16O}, 
we show the 
comparison of the calculated with the experimental sprectrum. In view of the 
fact that the states are complicated cluster states, the agreement between 
experiment and theory is very good. The interpretation goes as follows: the 
first four excited $0^+$ states have a $^{12}$C + $\alpha$ structure where 
the $\alpha$ orbits in 0S, 0D, 1S waves around the ground state 
core of $^{12}$C and 
also in a 0P wave orbiting around the first $1^-$ of $^{12}$C. 
It is only the first 
state above the four $\alpha$ particle threshold at 15.1 MeV  which is 
interpreted as an $\alpha$ gas state or a condensate. This state has, indeed, 
some analogies with the Hoyle state: it is just some hundreds of keV above 
threshold, it is strongly excited by inelastic electron scattering what means 
that monopole transition is quite large. Unfortunately the inelastic form 
factor has not been measured so far.


















\begin{figure}[h]
\hspace{0.7cm}
\begin{minipage}[h]{6.5cm}
\includegraphics[width=6.cm]{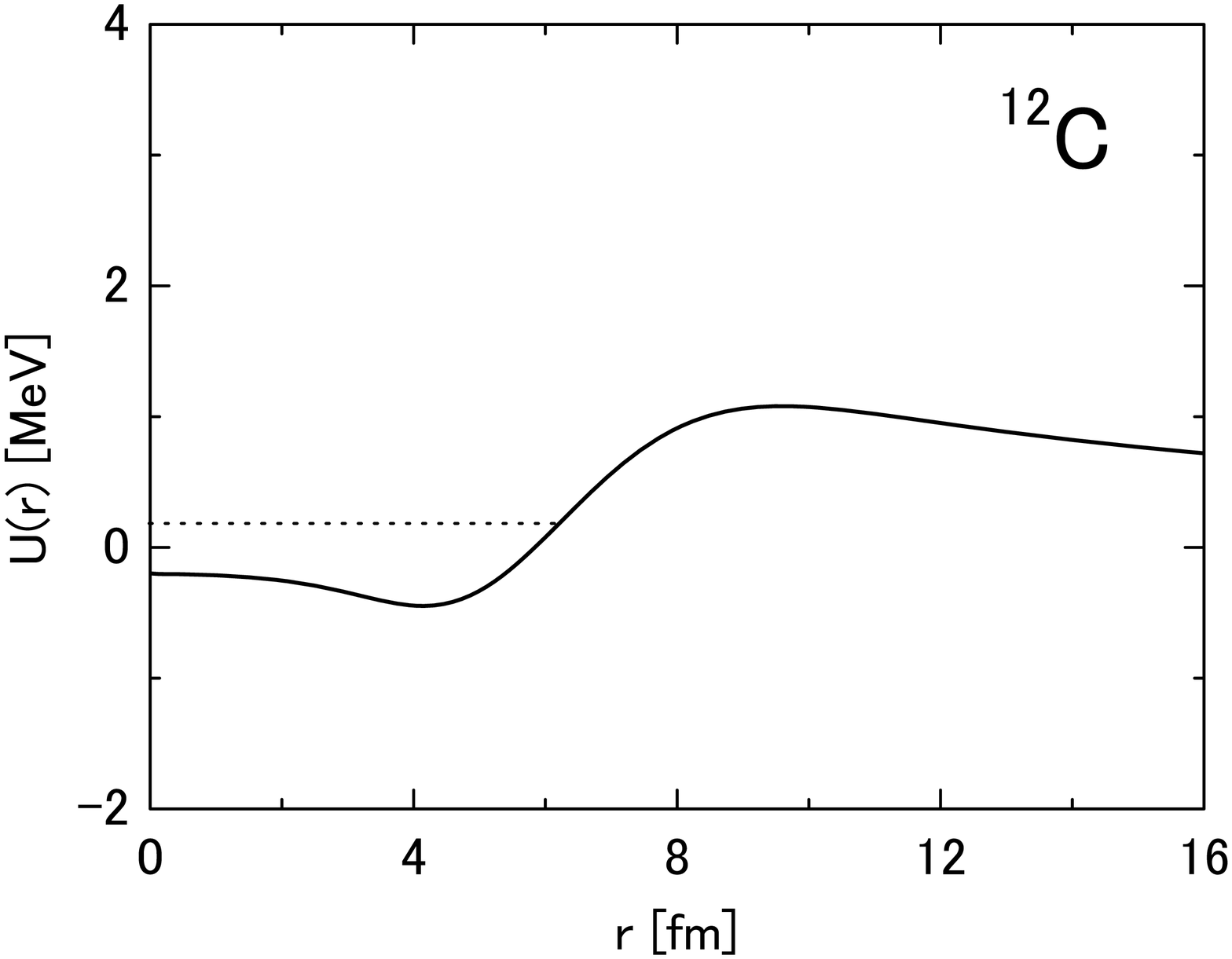}
\end{minipage}
\hfill
\begin{minipage}[h]{6.5cm}
\includegraphics[width=6.cm]{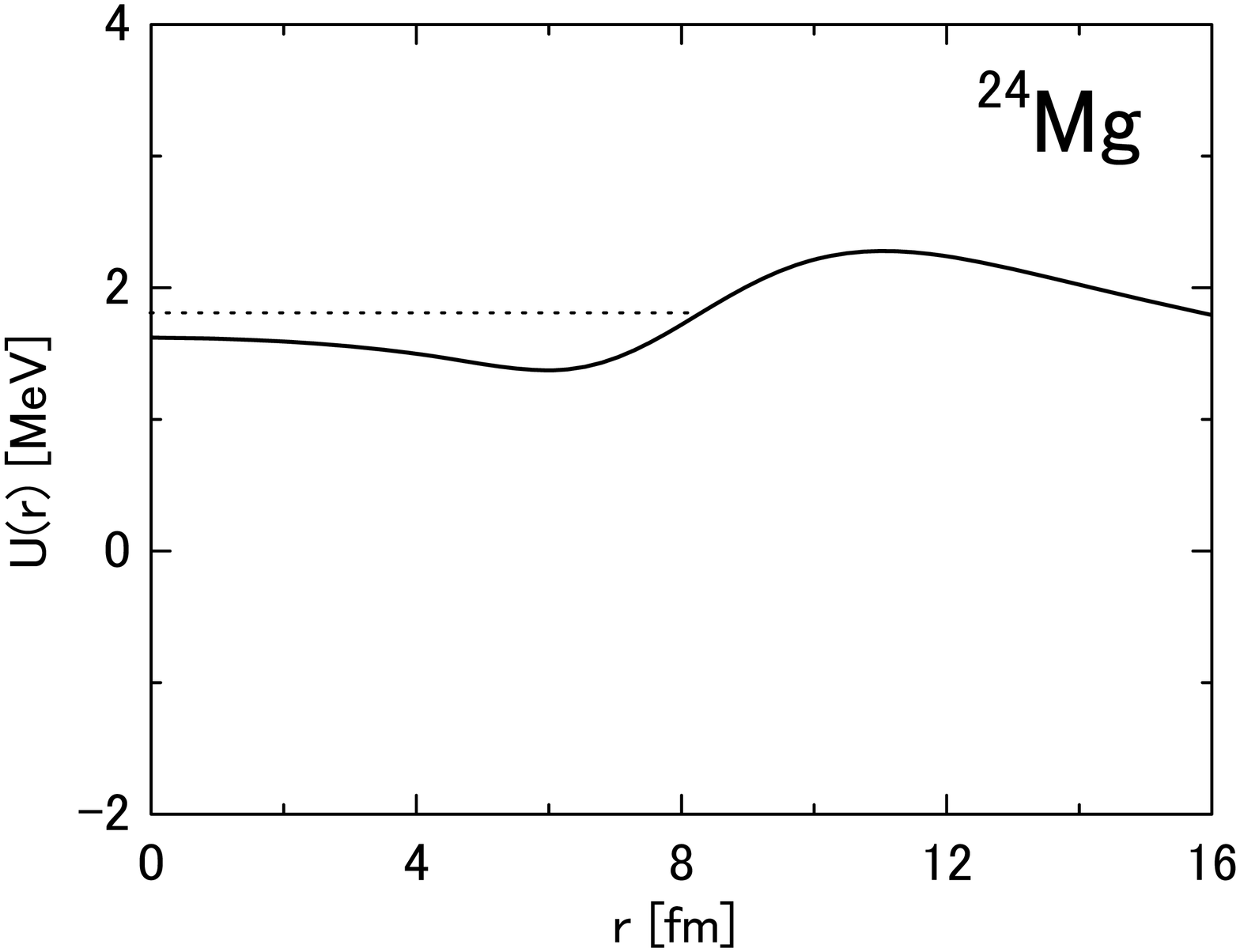}
\label{fig:4}
\end{minipage}
\caption{Alpha-particle mean-field potential for three $\alpha$'s in $^{12}$C and six $\alpha$'s in $^{24}$Mg. Note the lower Coulomb barrier for $^{24}$Mg (from Ref.~\cite{yamada2}).}
\end{figure}

One interesting question that can be asked at this point is: 
How many $\alpha$'s can maximally exist in a self-bound 
$\alpha$-gas state?  Seeking an answer,
we performed a schematic investigation using an effective 
$\alpha$-$\alpha$ interaction of the Ali-Bodmer form \cite{ali} within 
an $\alpha$-gas mean-field 
calculation of the Gross-Pitaevskii type \cite{gross}. 
The parameters of the force were slightly adjusted to reproduce 
our microscopic results for $^{12}$C. The corresponding 
$\alpha$ mean-field potential is shown in Fig.~\ref{fig:4}. 
One sees the $0S$-state lying slightly above threshold 
but below the Coulomb barrier.  As more $\alpha$-particles
are added, the Coulomb repulsion drives the loosely bound 
system of $\alpha$-particles farther and farther apart, so that the 
Coulomb barrier fades away.  According to our estimate \cite{yamada2}, 
a maximum of eight to ten $\alpha$-particles can be held together in a 
condensate.  However, there may be ways to lend additional stability
to such systems.  We know that in the case of $^8$Be, adding one or two 
neutrons produces extra binding without seriously disturbing the pronounced 
$\alpha$-cluster structure.  Therefore, one has reason to speculate 
that adding a few of neutrons to a many-$\alpha$ state may 
stabilize the condensate.  But again, state-of-the-art microscopic
investigations are necessary before anything definite can be said 
about how extra neutrons will influence an $\alpha$-particle 
condensate.

Another interesting idea concerning $\alpha$-particle condensates was 
put forward by von Oertzen and collaborators \cite{koka,oertzen}.  Adding 
more and more $\alpha$-particles to 
the, e.g., $^{40}$Ca core,  
one will arrive sooner or later at the point of $\alpha$-particle drip.  
Therefore minimal further excitation may be sufficient to shake loose 
some $\alpha$-particles, so that an $n\,\alpha$-condensate 
could be created on top of an inert $^{40}$Ca core.
Similar ideas also have been advanced by Ogloblin \cite{ogloblin}, 
who envisions a three-$\alpha$-particle condensate on top 
of $^{100}$Sn, and earlier by Brenner and Gridnev, who have presented 
evidence of experimental detection of gaseous $\alpha$-particles 
in $^{28}$Si and $^{32}$S on top of an inert $^{16}$O core 
\cite{brenner}.\\

Interesting new theoretical developments generalizing the THSR wave function \cite{thsr} are presently going on, putting nuclear cluster physics on a completely new basis \cite{bo-ren}. For example, it was found that the rotational parity doublet ground state bands in $^{20}$Ne can be described with a slightly generalised THSR wave function. Also the intriguing question about the intrinsic cluster structure inherent to the THSR wave function is further elucidated in \cite{bo-ren}. In this respect it is also interesting that Hartree-Fock-Bogoliubov calculations for expanding $n\alpha$ nuclei show clusterisation into geometrical arrangements of the $\alpha$'s, as, e.g., a tetrahedron for $^{16}$O \cite{girod}.












\section{Conclusions, Discussion, Outlook}

We have investigated the role that pairing
and multiparticle correlations may play in nuclear matter existing 
in dense astrophysical objects and in finite nuclei.
A complete and quantitative description of nuclear matter
must allow for the presence of clusters of nucleons, 
bound or metastable, possibly forming a quantum condensate.
In particular, quartetting correlations, responsible 
for the emergence of $\alpha$-like clusters, are identified
as uniquely important in determining the behavior of nuclear 
matter in the limiting regime of low density and low temperature.
We have calculated the transition temperature for the onset 
of quantum condensates made up of $\alpha$-like and deuteron-like 
bosonic clusters, and considered in considerable detail the 
intriguing example of Bose-Einstein condensation of $\alpha$ 
particles. It turns out that contrary to pairing, quartet condensation 
primarily exists in the BEC phase at low density. In which way quartet 
condensation is lost by increasing the density is still an open question. 
It may be similar to a liquid-gas phase transition. 
Anyway, it is clear that there can not exist a condensate of quartets with a 
long coherence length for arbitrarily small attraction as this is the case 
for pairing in the BCS phase. Arguments for this based on many 
fermion level densities have been presented.
It is inevitable that under increasing density or pressure,
the bound $\alpha$, $d$, or other nuclidic clusters, 
present at low density, experience significant modification due
to the background medium (and eventually merge with it).  
We have shown how self-energy corrections and Pauli blocking 
alter the properties of cluster states, 
and we have formulated a cluster mean-field approximation to 
provide an initial description of this process.  One result
of special interest is the suppression of the $\alpha$-like 
condensate, which is dominant at lower densities, as the density 
reaches and exceeds the Mott value, allowing the pairing 
transition to occur. Even at lower densities $\alpha$-particle condensation 
may be influenced by neutron excess, i.e. in the case of asymmetric nuclear 
matter, see Fig. 3. A genuine theory for the quartet order parameter in 
homogeneous infinite matter, similar to BCS theory, is demanded. First results 
in this direction have been published \cite{Sogo3}. A theory for quartet condensation on firmer grounds which parallels the one of pairing is presented in this contribution in section 2.

 A truly remarkable manifestation of $\alpha$-particle condensation 
seems to be present in finite nuclei. Indeed, the so-called Hoyle 
state (${0_2}^+$) in $^{12}$C at 7.654 MeV is very likely a 
dilute gas of three $\alpha$-particles, held together only by 
the Coulomb barrier.  This view is encouraged by the fact that we
can explain all the experimental data in terms of a conceptually 
simple wave function of the quartet-condensate type.  Within
the same model, we also systematically predict such states 
in heavier $n\,\alpha$ nuclei, and the search is on for their 
experimental identification. With the more phenomenological OCM method, we 
found that in $^{16}$O the sixth $0^+$ state at 15.1 MeV should be the 
candidate for an $\alpha$ condensed state. It is quite natural that such
states should exist up to some maximum number of $\alpha$
particles inspite of their increasing excitation energy: these $\alpha$-gas states are almost orhtogonal to the rest of nuclear states so that decay is strongly hindered.  We estimate that the phenomenon will terminate 
at about eight to ten $\alpha$'s as the confining Coulomb barrier 
fades away.  However, there is the possibility that larger 
condensates could be stabilized by addition of a few neutrons.  
Indeed, consider $^{9}$Be, which, contrary to $^{8}$Be, is bound by $\sim$ 
1.5 MeV, still showing a 
pronounced two $\alpha$-structure similar to the one of 
Fig.~\ref{profiles} (b). One could 
imagine ten $\alpha$'s or more, stabilised by two or four extra neutrons in a 
low density phase. However, even without being stabilised, if a compressed 
hot nuclear blob as e.g. produced in a central Heavy Ion collision expands and 
cools, it may turn on its way out, at a certain low density, into an 
expanding $\alpha$ condensed state 
where all $\alpha$'s are in relative S-waves. One may also, in one way or the 
other (photons?) excite, e.g., $^{40}$Ca to the 10 $\alpha$ threshold at 
about 60 MeV where a slow Coulomb explosion of an $\alpha$ gas would then take 
place. This would be an analogous 
situation to an expanding Bose condensate of atoms, once the trapping potential
has been switched off. Future dedicated experiments with high resolution 
multiparticle detectors  will tell whether such 
scenarios can be realised. Intriguing news in this respect come from GANIL 
where one may have achieved the disintegration of $^{56}$Ni into 14 $\alpha$'s 
(seven $\alpha$'s have been detected with high yield, the other seven may not 
have been seen because of the detectors were not sensitive to very low 
energy $\alpha$ particles \cite{Harakeh}\cite{Akimune}). Other possibilities of loose 
$\alpha$-gas states may exist on top of particularly stable cores, like 
$^{16}$O or $^{40}$Ca. Indeed in adding $\alpha$'s to e.g. $^{40}$Ca, one 
will reach the $\alpha$-particle drip line. Compound states of heavy N = Z 
nuclei of this kind may be produced in heavy ion reactions and an enhanced 
$\alpha$-decay rate may reveal the existence of an $\alpha$-particle 
condensate. Ideas of this type have been promoted by von Oertzen 
\cite{oertzen}, and also 
M. Brenner \cite{brenner}, and A. Ogloblin \cite{ogloblin}. However,  
coincidence measurements of 
multiple $\alpha$'s of decaying lighter nuclei like $^{16}$O may also be very 
useful \cite{freer}\cite{bord} \cite{zarub} to detect at least one additional 
$\alpha$-
condensate 
state 
beyond the only one that has been  identified so far, 
namely the ${0_2}^+$-state in $^{12}$C.

Another issue which may be raised in the context of $\alpha$-particle 
condensation is the question, also discussed in condensed matter physics 
\cite{noz-jam}, whether $\alpha$'s condense as singles or as doubles, i.e. 
as $^{8}$Be. In microscopic studies of $^{12}$C one, indeed, can see that in 
the ${0_2}^+$-state two of the three $\alpha$'s are slightly more closer to 
one another than to the third one \cite{feld}. The question is definitely very 
interesting and desserves future investigation, for instance in what concerns the identification of the $\alpha$ structure of the Hoyle state with the THSR wave function in the intrinsic frame. However, quantitatively, this
constitutes probably only a slight modification over the 
present formulation of $\alpha$-condensation. \\

Very recent triple $\alpha$ coincidence experiments \cite{3a-decay} have shown that there exists a decay channel of the Hoyle state where the three $\alpha$'s share democratically the available energy, that is, each $\alpha$ carries away one third of the initial energy. This very rare three body decay may give further credit to the idea that the three $\alpha$ particles occupy the same 0S-orbit , that is, they are condensed in this state. Same results have been confirmed by a second group \cite{3a-itoh} and earlier, with more yield, with heavy ion reactions \cite{borderie}.

What about 'ab initio' calculations of the Hoyle state? Presently several 
groups are on the track \cite{Meissner}\cite{Wiringa}. The Los Alamos group 
has achieved to calculate the density of the Hoyle 
state, though it is not yet converged in the far tail \cite{Wiringa}. 
It agrees very 
well with the density from the THSR wave function besides in the far tail.\\

In conclusion, we see that the idea of $\alpha$-particle 
condensation in nuclei and nuclear systems has triggered many new ideas 
, calculations, and experiments, in spite of the fact that, so far, 
a compelling 
case for such a state has only been made in $^{12}$C.  Even so,
the possible existence of a completely new nuclear phase in which
$\alpha$-particles play the role of quasi-elementary constituents is 
surely fascinating.  Hopefully, many more $\alpha$-particle states 
of nuclei will be detected in the near future, 
bringing deeper insights into the role of clustering and 
quantum condensates in systems of strongly interacting fermions.

Let us mention in the end that a more elaborate report on $\alpha$ particle 
condensation can be found in \cite{book}

\section{Acknowledgements}

We thank T. Sogo for contributions. Useful discussions and a careful reading of the mauscript by M. Tohyama are greatfully achknowledged. Discussions with B. Zhou and Z. Ren have been appreciated.


\begin{thebibliography} {99}
\bibitem{exstB}
The fact that this condensate state is actually an excited state of $^{12}$C 
does not invalidate this picture, since, on nuclear scales, this state is 
long lived. Actually, also atomic Bose condensates in traps find themselves 
in meta-stable states, see e.g. \cite{String}.
\end{thebibliography}
\end{document}